\documentclass{ifacconf}

 \usepackage{fancybox,amssymb,amsfonts,amstext,amsmath,stmaryrd,wasysym,mathtools,mathrsfs,cite}
\usepackage{graphicx} %for figure environment and include graphics command
\usepackage{natbib}
\usepackage{chemarrow}
\usepackage{circuitikz}
\usepackage{pgfplots,pgfplotstable}
\usepackage{mathtools}
\usepackage{url}
\usepackage[pdftex,all]{xy}
\SelectTips{cm}{}  % Tips (of arrows) are in accordance with Computer Modern

\newif\ifignore % when set to true, additional text appears containing
\ignorefalse
\newcommand{\auxproof}[1]{
\ifignore\mbox{}\newline
\textbf{PROOF:} \dotfill\newline
{\it #1}\mbox{}\newline
\textbf{ENDPROOF}\dotfill
\fi}

\newtheorem{mythm}{Theorem}[section]
\newtheorem{mylem}[mythm]{Lemma}

\newtheorem{mydef}[mythm]{Definition}
\newtheorem{myrem}[mythm]{Remark}
\newtheorem{myasm}[mythm]{Assumption}

 \def\myqed{\qed}

\newcommand{\N}{\mathbb{N}}
\newcommand{\R}{\mathbb{R}}

\newcommand{\Rplus}{\mathbb{R}^+}
\newcommand{\K}{{\mathcal K}}
\newcommand{\Signal}{{\mathcal S}}
\newcommand{\ON}{{\textit{ON}}}
\newcommand{\OFF}{{\textit{OFF}}}

\newcommand{\Pow}{{\mathcal P}}
\newcommand{\defeq}{\vcentcolon=}

\newcommand{\KL}{{\mathcal{KL}}}

\newcommand{\x}[3]{\mathbf{x}(#2, #1, #3)}

\newcommand{\Ra}[2]{\autorightarrow{\scriptsize\raisebox{-1.1ex}[0pt][0ex]{$#1$}}{\scriptsize\raisebox{1.1ex}[0pt][0pt]{$#2$}}}

\begin{document}
\begin{frontmatter}
 
\title{
Bounding Errors Due to Switching Delays in Incrementally Stable Switched Systems (Extended Version)\thanksref{thanks}}

\thanks[thanks]{Thanks are due to Georgios Fainekos, Krzysztof Czarnecki, Toshimitsu Ushio and the anonymous reviewers 
%on the occasion of submission to HSCC 2018
for the previous version
 for helpful discussions and comments. 
The authors are supported by JST ERATO HASUO Metamathematics for Systems Design Project
(No.\ JPMJER1603), and JSPS Grant-in-Aid No.\ 15K11984. 
 K.K.\ is supported by 
%JSPS under
% JSPS
 Grant-in-Aid for JSPS  Fellows No.\ 15J05580.}

\author[Kengo]{Kengo Kido}
\author[Sean]{Sean Sedwards}
\author[Ichiro]{Ichiro Hasuo}

\address[Kengo]{University of Tokyo \& JSPS Research Fellow, Tokyo, Japan}
\address[Sean]{University of Waterloo, Waterloo, ON, Canada}
\address[Ichiro]{National Institute of Informatics, Tokyo, Japan}


\begin{abstract}
{\em Time delays} pose an important challenge in networked control systems, which are now ubiquitous.
% Now that networked  control is ubiquitous,  \emph{time delays}  pose  a pressing challenge in control systems.
Focusing on switched systems, we introduce a
% approximate bisimulation-based
 framework that provides an upper bound for errors caused by switching delays. 
Our framework is based on \emph{approximate bisimulation}, a notion that has been previously utilized mainly for symbolic (discrete) abstraction of state spaces% \red{, initiated by Girard, Pola and Tabuada}
. Notable in our framework is that, in deriving  an approximate bisimulation and thus  an error bound, we use a %rather
simple incremental stability assumption (namely $\delta$-GUAS) that does not itself refer to time delays. That this is the same assumption used for  state-space discretization enables a \emph{two-step workflow} for control synthesis for switched systems, in which a single Lyapunov-type stability witness serves for two different purposes of state discretization and coping with time delays. 
We demonstrate the proposed framework with a boost DC-DC converter, a common example of switched systems.

\end{abstract}
\end{frontmatter}

\section{Introduction}
\label{sec:introduction}
% Thanks to the exponential growth in  capability and
%  availability of computer networks, 
\paragraph*{Time Delays}
\emph{Networked control}---digital control of physical systems via computer
networks---represents
%has become pervasive in modern control applications. Networked control
%represents
 an important aspect in various emerging system design paradigms, such as
cyber-physical systems and the Internet of Things. Consequently, identifying
and addressing challenges inherent in networked control has become a crucial part of the design of reliable real-world systems. 

The biggest challenge in networked control,
besides limited communication capacities and packet losses,
is \emph{time delay}. 
Physical separation of plants from controllers 
%by physical distances
 leads to inevitable communication delays dictated by the speed of light.  Worse, the rise of \emph{cloud control} is making both physical and logical distances 
between
% system
 components even longer and more unpredictable. 
%In such settings, 
Precise estimation of communication delays is often hard, let alone reducing them.

%kokokara!
These trends in control engineering call for a uniform
% control
 framework for robustness against potential time delays. In this paper, inspired by the
 hybrid nature of systems that is intrinsic to networked control,  we
 turn to \emph{approximate bisimulation} for coping with delays. 

 \paragraph*{Approximate Bisimulation}
An approximate bisimulation is a binary relation
between states of two systems, that witnesses the proximity of the systems'
behaviors. The notion  was first introduced in~\cite{Girard2007} as a
quantitative relaxation of \emph{bisimulation}, a well-established
coinductive equivalence notion between discrete transition systems
by~\cite{Park81}.
 Approximate bisimulation has  been
 actively studied ever since:
a notable theoretical result is
% The theory of approximate bisimulation has  been
% rapidly developed ever since; one  notable result is
its connection to
\emph{incremental stability} in~\cite{PolaGT08};
on the application side, it has been widely used in
 (discretized) \emph{symbolic abstraction} of continuous systems.
% \red{See e.g.~\cite{GirardPT10}.} 

In this paper we focus on switched systems, and use an approximate
 bisimulation for   error bounds between: a system
 $\Sigma_{\tau,\delta_{0}}$ with bounded time delays; and the corresponding system $\Sigma_{\tau}$ without delays. The choice of switched systems as our subject is justified by the envisaged applications in networked control. In a switched system, a plant has finitely many operation modes; and mode changes are dictated by 
 a switching signal that is sent from a controller over a computer network. This simple modeling encompasses various real-world networked control systems% \red{, as argued e.g.\ in~\cite{GirardPT10}}
 . Moreover, it turns out that our focus on switched systems greatly aids the analysis of effects of time delays.

 % One can use an approximate bisimulation for ensuring proximity between an actual system (with potential time delays) and an idealized system without delays. The latter system is simpler and one can use it for the purpose of control design; then the resulting controller is guaranteed to work well with the actual system, up-to certain errors that are bounded by the approximate bisimulation.

\paragraph*{Approximate Bisimulation for Switching Delays}
 Our  contributions in technical terms are as follows.
Our system model  $\Sigma_{\tau,\delta_{0}}$  is  a (potentially
nonlinear) switched system where switching signals are nearly periodic
with  a period $\tau$; the system exhibits potential switching delays
within a prescribed bound $\delta_{0}$. Our interest is in the
difference between the behaviors of $\Sigma_{\tau,\delta_{0}}$  and
those of the delay-free simplification $\Sigma_{\tau}$. We turn the two
systems into (discrete-time)  transition systems
$T(\Sigma_{\tau,\delta_{0}})$ and  $T(\Sigma_{\tau})$; between them 
we establish an approximate bisimulation that witnesses
 proximity of their behaviors. 
The approximate bisimulation is derived from
 an \emph{incremental stability} assumption of the dynamics of the
 system $\Sigma_{\tau,\delta_{0}}$
 (namely \emph{$\delta$-GUAS}). More specifically,
%  the
% incremental stability assumption as in~\cite{GirardPT10} (namely
% \emph{$\delta$-GUAS}).
%More specifically,
we present a construction that turns a Lyapunov-type witness for $\delta$-GUAS
%---namely common and multiple $\delta$-GAS Lyapunov functions---
into an approximate bisimulation.

Our workflow  just described resembles those in existing works
about the use of approximate bisimulation. That is,
\begin{itemize}
 \item starting from an
incrementally stable system $T$, one devises an abstraction
$T^{\text{abst}}$ of the system, 
 \item establishing an approximate
bisimulation between $T$ and $T^{\text{abst}}$ out of a Lyapunov-like
witness of stability.
 \item  Then the outcome of analysis of $T^{\text{abst}}$
(verification, control synthesis, etc.) can be carried over to the
original system $T$, modulo the error bounded by the approximate
bisimulation. 
\end{itemize}
One novelty of the current work is that, unlike most of
the existing works that aim at a symbolic (discrete) abstraction of a
state space, our abstraction $T^{\text{abst}}=\Sigma_{\tau}$ is a
\emph{temporally} idealized system without switching delays.

% This is not the first work that exploits Lyapunov-like witnesses of stability  for the purpose of bounding errors caused by time delays:  the works~\cite{PolaPBT10,PolaPB10} do so.
Lyapunov-like witnesses of stability, for the purpose of bounding errors
caused by time delays, have already been used in the works by~\cite{PolaPBT10,PolaPB10}.
Compared to these existing works, the current work is distinguished in
 that we rely only on a standard and relatively simple notion of stability, that does not
refer to time delays per se. Indeed, what we rely on is $\delta$-GUAS---the same
stability notion used for state-space discretization in many existing
works% \red{, such as~\cite{GirardPT10}}
.\footnote{We also identify some
additional technical constraints besides  $\delta$-GUAS (such as Assumption~\ref{asm:boundedIntermodeDerivative}) that are unique to the current setting.}
The abundance of analyses of
$\delta$-GUAS in the literature (e.g.~\cite{PolaGT08,GirardPT10,Girard10}) suggests that our use of it is an
advantage when it comes to application to various concrete systems.
See~\S{}\ref{sec:relatedWork} for further discussions on related work.

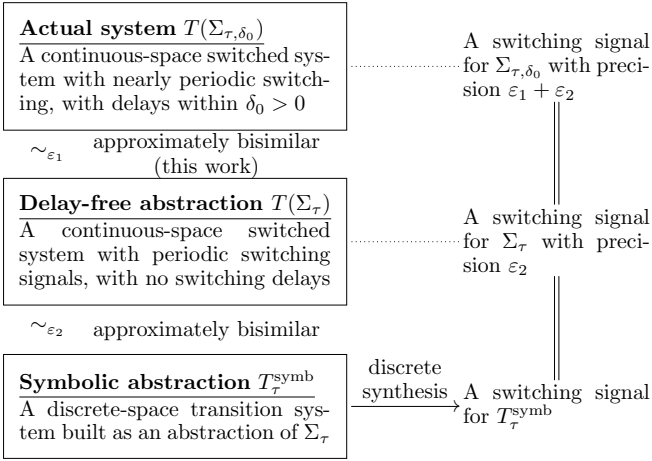
\begin{figure}[tbp]
\scalebox{.82}{
 \begin{math}  
  \def\labelstyle{\textstyle}
  \def\twocellstyle{\textstyle}
  \vcenter{\xymatrix@R=2em@C+2.5em{
  {\boxed{
   \begin{minipage}{5cm}
    \underline{{\bfseries Actual system} $T(\Sigma_{\tau,\delta_{0}})$}\\
 	  A
           continuous-space
          switched system 
 	  with nearly periodic switching, with delays  within $\delta_{0} >0$
    \end{minipage}
  }
  }
   \ar@{}[d]|-{\sim_{\varepsilon_{1}}\quad\txt{approximately bisimilar\\ (this work)}}
   \ar@{.}[r]
  &
  {{
   \begin{minipage}{3cm}
    A switching signal for $\Sigma_{\tau,\delta_{0}}$ with precision $\varepsilon_{1}+\varepsilon_{2}$
    \end{minipage}
  }}
  \\
  {\boxed{
   \begin{minipage}{5cm}
    \underline{{\bfseries Delay-free abstraction $T(\Sigma_{\tau})$}}\\
 	  A 
          continuous-space
          switched system with periodic switching signals, 
 	  with no switching delays
    \end{minipage}
  }}
   \ar@{}[d]|-{\sim_{\varepsilon_{2}}\quad\txt{approximately bisimilar%\\ \red{(by e.g.~\cite{GirardPT10})}
  }}
   \ar@{.}[r]
  &
  {{
   \begin{minipage}{3cm}
    A switching signal for $\Sigma_{\tau}$ with precision $\varepsilon_{2}$
    \end{minipage}
  }}
   \ar@{=}[u]
  \\
  {\boxed{
   \begin{minipage}{5cm}
    \underline{{\bfseries Symbolic abstraction} $T_{\tau}^{\text{symb}}$}\\
 	  A discrete-space transition system built as an abstraction of $\Sigma_{\tau}$
 % on which controllers can be synthesized based on techniques for DESs and algorithmic games
    \end{minipage}
  }}
  \ar[r]^-{\txt{discrete\\ synthesis}}
 &
  {{
   \begin{minipage}{3cm}
    A switching signal for $T_{\tau}^{\text{symb}}$ 
    \end{minipage}
  }}
   \ar@{=}[u]
}}
\end{math}
}
\caption{A two-step control synthesis workflow for switched systems with
 delays. We separate two concerns:  time delays and state-space
 discretization.  The same stability assumption on $\Sigma_{\tau}$ can
 be used once for all, for establishing both $\sim_{\varepsilon_{1}}$ and  $\sim_{\varepsilon_{2}}$.}
\label{fig:usageScenario}
\end{figure}
\paragraph*{Two-Step Control Synthesis for Switched Systems with
 Delays}
Even better, our reliance on the standard notion of $\delta$-GUAS enables
the following two-step synthesis workflow, where we combine the current results and those in~\cite{GirardPT10}.
% in view of the following usage scenario of the current results. 
See Fig.~\ref{fig:usageScenario}.
Our current results are used to derive the first error bound $\varepsilon_{1}$ between (the transition system $T(\Sigma_{\tau,\delta_{0}})$ derived from) the original system $\Sigma_{\tau,\delta_{0}}$, and 
 (the transition system $T(\Sigma_{\tau})$ derived from) the delay-free abstraction 
 $\Sigma_{\tau}$. 
The latter system $\Sigma_{\tau}$ is a delay-free periodic switched system, to
 which we can apply the state-space discretization technique
 in~\cite{GirardPT10}. 
 We thus construct a
 discretized symbolic model $T_{\tau}^{\text{symb}}$ and establish the
 second approximate bisimulation $\sim_{\varepsilon_{2}}$ in
 Fig.~\ref{fig:usageScenario}. The fact that our construction relies on
 the same  stability assumptions used in~\cite{GirardPT10} means
 the following: for establishing both of the approximate bisimulations
 $\sim_{\varepsilon_{1}}$ and  $\sim_{\varepsilon_{2}}$,
%in Fig.~\ref{fig:usageScenario}, 
 we can reuse the same ingredient (namely
 a $\delta$-GUAS Lyapunov function), instead of finding two different
 Lyapunov functions. 

Once we obtain a symbolic  model, 
%follow the workflow in Fig.~\ref{fig:usageScenario}, 
we can apply to it
various discrete techniques, such as automata-theoretic synthesis (as in~\cite{Vardi1995}),
%model checking~\cite{Baier2008}, 
supervisory control of discrete event systems (as in~\cite{RamadgeW87}), algorithmic game theory (as in~\cite{ArnoldVW03}), etc.
This is the horizontal arrow at the bottom of Fig.~\ref{fig:usageScenario}. The resulting controller (i.e.\  a switching signal, in the current setting)
 is then guaranteed, by the two approximate bisimulations, to work well with $\Sigma_{\tau}$ (with precision $\varepsilon_{2}$) and with $\Sigma_{\tau,\delta_{0}}$ (with precision $\varepsilon_{1}+\varepsilon_{2}$).\footnote{Here \emph{precision} means an upper bound for errors. } 

This way we ultimately derive a switching signal for the original system  $\Sigma_{\tau,\delta_{0}}$ whose precision 
 $\varepsilon_{1}+\varepsilon_{2}$ is guaranteed. The workflow in
 Fig.~\ref{fig:usageScenario} takes two steps that separate
 concerns (namely time delays and discretization of state spaces). While
 this two-step approach can potentially lead to loss of generality
 (especially in comparison with~\cite{PolaPB10},
 see~\S{\ref{sec:relatedWork}}), it seems to help coping with the
 problem's complexity. We demonstrate our workflow in~\S{\ref{sec:example}}, where we successfully derive a controller for
 a boost DC-DC converter example % (a common example\red{, studied e.g.\ in~\cite{BeccutiPM05,GirardPT10}})
 with additional switching delays.

\auxproof{The composition  $\varepsilon_{1}+\varepsilon_{2}$ of error bounds is additive because the controller synthesized following~\cite{GirardPT10} does not need sampling of the dynamics.}

\paragraph*{Contributions} Overall, our contributions are summarized as follows. We present a construction of an approximate bisimulation between a nearly-periodic switched systems and its (exactly) periodic approximation. This allows us to bound the difference between trajectories due to switching delays.  Thanks to our focus on switched systems we can use a common stability assumption (namely $\delta$-GUAS) as the ingredient of the construction; this allows us to combine the current results with the existing results on  symbolic abstraction and control synthesis, leading to a two-step control synthesis workflow (Fig.~\ref{fig:usageScenario}) where the same stability analysis derives two approximate bisimulations. 

We defer proofs and another example (besides the boost DC-DC converter in~\S{}\ref{sec:example}) to the appendices.

% Cyber-physical systems
% Delay
% Network control
% Nonlinear
% Approximate bisimulation

\section{Switched Systems}
\label{sec:switchedSys}
The set of nonnegative real numbers is denoted by $\Rplus$. 
%One main usage of it is as the domain of time. 
We let
  $\|\_\|$ denote the usual Euclidean norm on $\R^{n}$. 
\begin{mydef}[switched system]\label{def:switchedSystem}
 A \emph{switched system} is a quadruple $\Sigma = (\R^n, P, \Pow, F)$ that consists of:
\begin{itemize}
 \item a \emph{state space} $\R^n$;
 \item a finite set $P = \{1, 2, \dotsc, m\}$ of \emph{modes};
 \item a set of \emph{switching signals} $\Pow \subseteq \Signal(\Rplus, P)$, where $\Signal(\Rplus, P)$ is the set of functions from $\Rplus$ to $P$ that satisfy the following conditions:  1) piecewise constant, 2) continuous from the right, and 3) non-Zeno;
% \item the set of switching signals $\Pow$ is a subset of $\Signal(\Rplus, P)$, where $\Signal(\Rplus, P)$ denotes the set of infinite or finite sequences of $(t_i, m_i)\in(\Rplus \times P)$ such that the sequence $(t_i)_{i\in\N}$ is strictly increasing, starting from $t_0 = 0$ and non-Zeno;
 \item $F = \{f_1, f_2, \dotsc, f_m\}$ is the set of vector fields indexed by $p\in P$, where each  $f_p$ is a locally Lipschitz continuous function from $\R^n$ to $\R^n$.
\end{itemize}
\end{mydef}

 A continuous and piecewise $\mathcal{C}^1$ function $\mathbf{x}: \Rplus\rightarrow\R^n$ is called  a \emph{trajectory} of the switched system $\Sigma$ if there exists a switching signal $\mathbf{p}\in\Pow$ such that $\dot{\mathbf{x}}(t) = f_{\mathbf{p}(t)}(x(t))$
 holds at each time $t\in\Rplus$ when the switching signal $\mathbf{p}$ is continuous.

We let  $\x{x}{t}{\mathbf{p}}$ denote the point reached at time $t\in\Rplus$, starting from the state $x\in\R^n$ (at $t=0$), under the switching signal $\mathbf{p}\in\Pow$.
In the special case where the switching signal is constant (i.e.\ $\mathbf{p}(s)=p$ for all $s\in\Rplus$),it is denoted by 
 $\x{x}{t}{p}$.
The continuous subsystem of $\Sigma$ with the constant switching signal $\mathbf{p}(s)=p$ for all $s\in\Rplus$ is denoted by $\Sigma_p$.
If $P$ is a singleton $P=\{p\}$, the system $\Sigma = \Sigma_p$ is a continuous system without switching.

\begin{figure}[tbp]
 \noindent\hspace{-1em}
  \includegraphics{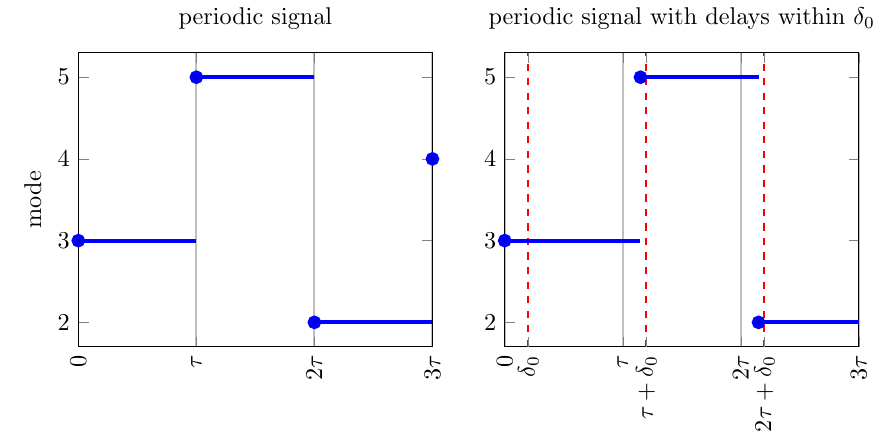}
  \caption{Periodic switching signals, with and without delays.}
  \label{fig:withAndWithoutDelays}
 \end{figure}

\begin{mydef}[periodicity, switching delay]\label{def:switchingDelay}
Let $0\leq\delta_{0}<\tau$.  A switching signal $\mathbf{p}$ is said to be \emph{$\tau$-periodic with switching delays within $\delta_{0}$}
if it has at most one discontinuity in each interval $[k\tau, k\tau+\delta_{0}]$  (where $k\in\N$), and constant elsewhere.
The time instants $t\in\Rplus$ where $\mathbf{p}$ is discontinuous are called \emph{switching times}.
A switched system $\Sigma = (\R^n, P, \Pow, F)$ is called \emph{$\tau$-periodic with switching delays within $\delta_{0}$} if all the switching signals in $\Pow$ are $\tau$-periodic with switching delays within $\delta_{0}$.
Additionally, if $\delta_0 = 0$, a switching signal or a switched system is called \emph{$\tau$-periodic}.
% Given a switching signal $\mathbf{p}$, 
% If a switching signal is continuous except at $k\tau$ (where $\tau>0$ is a constant and $k\in\N$), it is called
% A switched system is called \emph{$\tau$-periodic} if all the switching signals in $\Pow$ are $\tau$-periodic.
\end{mydef}
See Fig.~\ref{fig:withAndWithoutDelays} for illustration of periodic switching signals and those with delays. 

\begin{myrem}
Our results rely on $\delta_0<\tau$, which we believe is a reasonable assumption.
For example, in automotive applications, common switching periods are 4--8 milliseconds while jitters arising from CAN (Controller Area Networking) latency can be bounded by 120 microseconds.
\end{myrem}

In this paper we focus on periodic switched systems with switching delays, and their difference from those without switching delays. More specifically, 
we consider two switched systems
\begin{equation}\label{eq:twoSwitchedSys}\small
\begin{aligned}
    \Sigma_{\tau, \delta_{0}} &= (\R^n, P, \Pow_{\tau, \delta_{0}}, F)
  &&\text{$\tau$-periodic with delays  $\leq\delta_{0}$}
  \\
  \Sigma_\tau &= (\R^n, P, \Pow_{\tau}, F)
  &&\text{$\tau$-periodic}
\end{aligned}
\end{equation}
%$\Sigma_{\tau, \delta_{0}} = (\R^n, P, \Pow_{\tau, \delta_{0}}, F)$ and $\Sigma_\tau = (\R^n, P, \Pow_{\tau}, F)$ 
that have $\R^{n}$, $P$ and $F$ in common.
For the former system $\Sigma_{\tau, \delta_{0}}$,
%$\Sigma_{\tau, \delta_{0}}$ is $\tau$-periodic with switching delays within $\delta_{0}$ (meaning that
 the set $\Pow_{\tau, \delta_{0}}$  consists of all $\tau$-periodic signals with  delays within $\delta_{0}$; for
 % meaning that  the set $\Pow_{\tau, \delta_{0}}$ of switching signals consists of all $\tau$-periodic with actuation delays $\delta_{0}$ switching signals.
 the latter system $\Sigma_\tau$
% is a $\tau$-periodic switched system (meaning that
 the set $\Pow_{\tau}$ consists of all $\tau$-periodic switching signals.
% Given a switching signal $p\in \mathcal{P},$ we consider the sequence $\pi_p = ((t_0, m_0), (t_1, m_1),  (t_2, m_2), \cdots) (0=t_0\leq t_1 \leq t_2 \leq, \cdots)$ consisting of $t_i\in\Rplus (i\in\N)$ such that $p$ is not continuous at $t_i$.

\section{Transition Systems and Approximate Bisimulation}\label{sec:transSys}
 We use approximate bisimulations from~\cite{Girard2007} to formalize  proximity between  $\Sigma_{\tau, \delta_{0}}$ (with delay) and $\Sigma_{\tau}$ (without). In this section we present our key definition (Def.~\ref{def:TSigma}) that allows such use of approximate bisimulation, in addition to  a quick recap of a basic theory of approximate bisimulation.

\begin{mydef}[transition system]\label{def:transitionSystems}
 A \emph{transition system} is a triple $T = (Q, L, \Ra{}{}, O, H, I)$,
% consisting of a set $Q$ of states, a set $L$ of labels, a transition relation $\Ra{}{} \subseteq Q\times L\times Q$, a set $O$ of outputs, an output function $H: Q \rightarrow O$ and a set $I$ of initial states.
 where
\begin{itemize}
 \item $Q$ is a set of states;
 \item $L$ is a set of labels;
 \item $\Ra{}{} \subseteq Q\times L\times Q$ is a transition relation;
 \item $O$ is a set of outputs;
 \item $H: Q \rightarrow O$ is an output function; and
 \item $I\subseteq Q$ is a set of initial states.
\end{itemize}
We let  $q\Ra{l}{}q'$ denote the fact that $(q, l, q')\in \Ra{}{}$.

In this paper, for a set $X$,  a function $d: X\times X \rightarrow \Rplus \cup \{\infty\}$ that satisfies, for all $x, y, z\in X$, $d(x, y)\geq 0$ % , $d(x, x) = 0$,
 and $d(x, z)\leq d(x, y)+d(y, z)$ is called a \emph{premetric} on $X$.
A transition system $T$ is said to be \emph{premetric} if the set $O$ of outputs is equipped with a premetric $d$.
\end{mydef}

 Approximate bisimulations are defined between transitions systems. 
It is a (co)inductive construct that guarantees henceforth proximity of behaviors of two states.
\begin{mydef}
 Let $T_i = (Q_i, L, \Ra{}{i}, O, H_i, I_i)$ ($i=1$, $2$) be two
 premetric transition systems,
sharing the same sets of actions $L$ and outputs $O$ with a premetric $d$.
Let $\varepsilon\in\Rplus$ be a positive number; we call it a \emph{precision}. 
%Given a \emph{precision} $\varepsilon\in\Rplus$, 
A relation $R\subseteq
 Q_1\times Q_2$ is called an \emph{$\varepsilon$-approximate bisimulation relation} between $T_1$ and $T_2$ if the following three conditions hold for all $(q_1, q_2)\in R$.
\begin{itemize}
 \item $d(H_1(q_1), H_2(q_2))\leq\varepsilon$;
 \item $\forall q_1\Ra{l}{1}q'_1, \exists q_2\Ra{l}{2}q'_2 \text{ such that } (q'_1, q'_2)\in R$; and
 \item $\forall q_2\Ra{l}{2}q'_2, \exists q_1\Ra{l}{1}q'_1 \text{ such that } (q'_1, q'_2)\in R$.
\end{itemize}
The transition systems $T_1$ and $T_2$ are  \emph{approximately
 bisimilar with precision $\varepsilon$} if there exists an
 $\varepsilon$-approximate bisimulation relation $R$ that satisfies the
 following conditions:
\begin{itemize}
 \item $\forall q_1\in I_1, \exists q_2\in I_2 \text{ such that } (q_1, q_2)\in R$;
 \item $\forall q_2\in I_2, \exists q_1\in I_1 \text{ such that } (q_1, q_2)\in R$.
\end{itemize}
We let $T_1\sim_\varepsilon T_2$ denote the fact that $T_1$ and $T_2$ are approximately bisimilar with precision $\varepsilon$.
\end{mydef}

%sean
%For the  two switched systems $\Sigma_{\tau, \delta_{0}} = (\R^n, P, \Pow_{\tau, \delta_{0}}, F)$ and $\Sigma_{\tau} = (\R^n, P, \Pow_{\tau}, F)$ in~(\ref{eq:twoSwitchedSys}), we shall construct associated transition systems $T(\Sigma_{\tau, \delta_{0}})$ and $T(\Sigma_{\tau})$ respectively.
For the  two switched systems $\Sigma_{\tau, \delta_{0}} = (\R^n, P, \Pow_{\tau, \delta_{0}}, F)$ and $\Sigma_{\tau} = (\R^n, P, \Pow_{\tau}, F)$ in~(\ref{eq:twoSwitchedSys}), we shall construct associated transition systems $T(\Sigma_{\tau, \delta_{0}})$ and $T(\Sigma_{\tau})$, respectively.

% $$T(\Sigma_{\tau, \delta_{0}}) = (Q_{\tau, \delta_{0}}, L, \Ra{}{\tau, \delta_{0}}, O, H_{\tau, \delta_{0}}, I)$$ and
% $$T(\Sigma_{\tau}) = (Q_{\tau}, L, \Ra{}{\tau}, O, H_{\tau}, I)$$ in the following manner.

\begin{mydef}[$T(\Sigma_{\tau, \delta_{0}}), T(\Sigma_{\tau})$]
\label{def:TSigma}
The transition system
$$T(\Sigma_{\tau, \delta_{0}}) = (Q_{\tau, \delta_{0}}, L, \Ra{}{\tau, \delta_{0}}, O, H_{\tau, \delta_{0}}, I) \enspace,$$
% \begin{equation*}
%  \begin{aligned}
%   T(\Sigma_{\tau, \delta_{0}}) &= (Q_{\tau, \delta_{0}}, L, \Ra{}{\tau, \delta_{0}}, O, H_{\tau, \delta_{0}}, I) \enspace,
% % \qquad\text{and}
% %   \\
% %   T(\Sigma_{\tau}) &= (Q_{\tau}, L, \Ra{}{\tau}, O, H_{\tau}, I)
%  \end{aligned}
% \end{equation*}
associated with the switched system $\Sigma_{\tau, \delta_{0}}$ with delays in~(\ref{eq:twoSwitchedSys}), is defined as follows:
%The transition system $T(\Sigma_{\tau, \delta_{0}})$ is defined by:
\begin{itemize}
 \item the set of states is $Q_{\tau,\delta_{0}} \defeq \R^n\times \bigcup_{k\in\N}[k\tau, k\tau+\delta_{0}] \times P$;
 \item the set of labels $L$ is the set of modes, i.e.\ $L\defeq P$;
 \item the transition relation $\Ra{}{\tau, \delta_{0}}\subseteq Q_{\tau, \delta_{0}}\times L\times Q_{\tau, \delta_{0}}$ is defined by $(x, t, p) \Ra{p''}{\tau, \delta_{0}}(x', t', p')$ if $p=p''$, $x' = \x{x}{t'-t}{p}$ and there exists $k\in\N$ such that $t \in [k\tau, k\tau+\delta_{0}]$ and $t'\in[(k+1)\tau, (k+1)\tau+\delta_{0}]$;
 \item the set of outputs is $O\defeq \R^n\times\Rplus\times P$;
 \item the output function $H\colon Q_{\tau,\delta_{0}}\to O$ is the canonical embedding function $\R^n\times \bigcup_{k\in\N}[k\tau, k\tau+\delta_{0}] \times P\to \R^n\times\Rplus\times P$; and
 \item the set of initial states is $I\defeq \R^n\times\{0\}\times P$.
\end{itemize} 
Intuitively, each state $(x,t,p)$ of $T(\Sigma_{\tau,\delta_{0}})$ marks switching in the system $\Sigma_{\tau,\delta_{0}}$: $x\in \R^{n}$ is the (continuous) state at switching; $t$ is time of switching; and $p$ is the next mode. Note that, by the assumption on $\Sigma_{\tau,\delta_{0}}$, $t$ necessarily belongs to the interval $[k\tau, k\tau+\delta_{0}]$ for some $k\in\N$. 
\end{mydef}

The transition system
\begin{equation*}
 \begin{aligned}
%   T(\Sigma_{\tau, \delta_{0}}) &= (Q_{\tau, \delta_{0}}, L, \Ra{}{\tau, \delta_{0}}, O, H_{\tau, \delta_{0}}, I) \enspace,
% % \qquad\text{and}
%   \\
   T(\Sigma_{\tau}) &= (Q_{\tau}, L, \Ra{}{\tau}, O, H_{\tau}, I)\enspace,
 \end{aligned}
\end{equation*}
associated with the switched system $\Sigma_{\tau}$ without delays in~(\ref{eq:twoSwitchedSys}), is defined similarly, by fixing $\delta_0$ in the above definition to $0$.
% \begin{itemize}
%  \item the set of states is
% $Q_{\tau} \defeq \R^n\times \{0,\tau,2\tau,\dotsc\} \times P$;
% %$$\qquad Q_{\tau} \defeq \R^n\times \{t\in\Rplus\mid t = k\tau \text{ for some }k\in\N\} \times P;$$
%  \item the set of labels $L$ is the set of modes, i.e.\ $L\defeq P$;
%  \item the transition relation $\Ra{}{\tau}\subseteq Q_{\tau}\times L\times Q_{\tau}$ is defined by $(x, t, p) \Ra{p''}{\tau}(x', t', p')$ if $p=p''$, $t'=t+\tau$ and $x' = \x{x}{\tau}{p}$;
%  \item the set of outputs is $O\defeq \R^n\times\Rplus\times P$;
%  \item the output function $H\colon Q_{\tau,\delta_{0}}\to O$ is the canonical embedding function; and
% % \item the output function $H$ is the canonical embedding function; and
%  \item the set of initial states is $I\defeq \R^n\times\{0\}\times P$.
% \end{itemize} 

 Note that, in both of $T(\Sigma_{\tau,\delta_{0}})$ and $T(\Sigma_{\tau})$,
 the label $p''$ for a transition is uniquely determined by the mode
 component $p$ of the transition's source $(x,t,p)$. Therefore,
 mathematically speaking, we do not need transition labels.
%; they remain
 % in our formalization in order to provide intuitions.

In~\cite{GirardPT10}, the state space $Q$
of the transition system is just the continuous state space $\R^{n}$ of the switched system.
In comparison, ours has two additional components, namely time $t$ and the current mode $p$.
It is notable that moving a mode $p$ from transition labels to state labels allows us to analyze what happens during switching delays, that is, when the system keeps operating under the mode $p$ while it is not supposed to do so.

%sean
% In the work~\cite{GirardPT10} that we are based on, the state space $Q$
% of the transition system is defined to be $\R^{n}$ and is the same as the state space of the original switched system. In comparison our definition has two additional components, namely time $t$ and the current mode $p$. It is notable that moving a mode $p$ from transition labels to state labels allow us to analyze what can happen during switching delays, that is, when the system keeps operating under the mode $p$ while it is not supposed to do so.

% This construction of the transition systems $T(\Sigma_{\tau, \delta_{0}})$ and $T(\Sigma_{\tau})$ from the given switched systems is different from the one used in \cite{GirardPT10}.
% Our construction focuses on the times when a switching occurs.
% It is easy, however, to see our $T(\Sigma)$ simulates the transition system defined in \cite{GirardPT10}.

%Given $\delta_{0}$, 

\begin{mydef}[premetric on outputs]\label{def:pseudoquasimetricBetweenTrajectories}
The transition systems  $T(\Sigma_{\tau,\delta_{0}})$ and $T(\Sigma_{\tau})$ are premetric 
with the following $d$, defined on the common set of outputs $O=\R^n \times \Rplus \times P$:
% On the set of outputs $O=\R^n \times \Rplus \times P$ that is common to
% the two  transition systems  $T(\Sigma_{\tau,\delta_{0}})$ and $T(\Sigma_{\tau})$,  
% we define the following premetric $d$:
\begin{align*}
& d((x, t, p),(x', t', p'))\defeq\\
&\left\{
\begin{array}{ll}
 \| x - \x{x'}{t-t'}{p}\|\quad &
\begin{array}{ll} 
 \text{if }p = p', t'=k\tau \text{ and }\\
 \quad t\in[t', t'+\delta_{0}] \text{ for some } k\in\N
\end{array}\\
 \infty & \text{\ otherwise.}
\end{array}
\right.
\end{align*}
%Here  $\|\_\|$ denotes the usual Euclidean metric on $\R^{n}$. 
\end{mydef}

\section{Incremental Stability}\label{sec:incrStability}
After
the pioneering work by~\cite{PolaGT08},   
a number of frameworks rely on the assumption of
\emph{incremental stability} for the construction of approximate
bisimulations. Intuitively, a dynamical system is incrementally stable
if, under any choice of an initial state, the resulting trajectory
asymptotically converges to one reference trajectory. 
In this section, we review an incremental stability for switched systems, following \cite{GirardPT10}.

%sean
%In the following definitions we will be using the following classes of
In the subsequent definitions we will be using the following classes of
functions.
A continuous function $\gamma:\Rplus\rightarrow\Rplus$ is a \emph{class $\K$ function} if it is strictly increasing and $\gamma(0) = 0$.
A $\K$ function is a \emph{$\K_\infty$ function} if $\gamma(x)\rightarrow\infty$ when $x\rightarrow\infty$.
A continuous function $\beta:\Rplus\times\Rplus\rightarrow\Rplus$ is a
\emph{class $\KL$ function} if 1) the function defined by
$x\mapsto\beta(x, t)$ is a $\K_\infty$ function for any fixed $t$; and
2) for any fixed $x$, the function defined by $t\mapsto\beta(x, t)$ is strictly decreasing, and $\beta(x, t)\rightarrow 0$ when $t\rightarrow\infty$.

% % The notion of incremental stability means that if the system is started at two different states, their states is getting closer with time progress.
% % It is originally for continuous systems.
% % In the definitions, the following classes of functions will be used.

% \begin{mydef}[$\delta$-GAS system~\cite{DBLP:journals/tac/Angeli02}]\label{def:deltaGAS}
%  Let $\Sigma = (\R^n, P, \Pow, F)$ be a single-mode
%  switched system where $P=\{p\}$ is a
%  singleton  (therefore there is actually no switching). 
% The system $\Sigma$ is \emph{incrementally globally asymptotically stable ($\delta$-GAS)} if there exists a $\KL$ function $\beta$ such that $$\|\x{x}{t}{p} - \x{y}{t}{p}\|\leq\beta(\|x-y\|, t)$$
% for all $x, y \in \R^n$ and $t\in\Rplus$.
% \end{mydef}

% The notion of $\delta$-GAS is a well-known one among various notions of
% incremental stability. 

% The notions so far are for  systems without switching. 
% \red{Their extension
% %of incremental stability \cite{DBLP:journals/tac/Angeli02}
% to switched systems are introduced in~\cite{GirardPT10}.} 

 \begin{mydef}%[\cite{GirardPT10}]
  \label{def:deltaGUAS}
 Let $\Sigma = (\R^n, P, \Pow, F)$ be a switched system. $\Sigma$ is
 said to be \emph{incrementally globally uniformly asymptotically stable
 ($\delta$-GUAS)} if there exists a $\KL$ function $\beta$ such that the
 following holds
for all $x, y \in \R^n$, $t\in\Rplus$ and $\mathbf{p}\in\Pow$.
 $$\|\x{x}{t}{\mathbf{p}} - \x{y}{t}{\mathbf{p}}\|\leq\beta(\|x-y\|, t)$$
\end{mydef}

 Directly establishing that a system is $\delta$-GUAS is often hard. A usual technique in the field is to let a
Lyapunov-type function  play the role of
witness for $\delta$-GUAS. %~\cite{DBLP:journals/tac/Angeli02}. 

% This incremental stability is characterized by the existence of a variant of Lyapunov function.

\begin{mydef}\label{def:deltaGASLyapunov}
 Let $\Sigma = (\R^n, P, \Pow, F)$ be a single-mode switched system
 with $P=\{p\}$. A smooth function $V: \R^n \times \R^n \rightarrow
 \Rplus$ is a \emph{$\delta$-GAS Lyapunov function for $\Sigma$} if
 there exist $\K_\infty$ functions $\underline{\alpha}$,
 $\overline{\alpha}$ and $\kappa>0$ such that the following hold
 for all $x, y \in \R^n$.
\begin{align}
& \underline{\alpha}(\|x-y\|)\leq V(x, y) \leq \overline{\alpha}(\|x-y\|)\label{eq:defValpha}\\
& \frac{\partial V}{\partial x}(x, y)f_p(x) +
 \frac{\partial V}{\partial y}(x, y)f_p(y) \leq -\kappa V(x, y)\label{eq:defVkappa}
\end{align}
\end{mydef}
Note that the left-hand side of~(\ref{eq:defVkappa}) is much like the Lie derivative of $V$ along the vector field $f_{p}$. 

% \begin{mythm}[\cite{DBLP:journals/tac/Angeli02}]\label{thm:deltaGAS}
%  Let $\Sigma = (\R^n, P, \Pow, F)$ be a single-mode switched system
%  with $P=\{p\}$.
%  %Let $\Sigma = (\R^n, P=\{p\}, \Pow, F)$ be a continuous system.
% The system $\Sigma$ is $\delta$-GAS if and only if it has a $\delta$-GAS Lyapunov function.
% \qed
% \end{mythm}

A sufficient condition for a switched system to be $\delta$-GUAS is the
existence of a common $\delta$-GAS Lyapunov function. 

%  \begin{mydef}%[\cite{GirardPT10}]
%   \label{def:commondeltaGASLyapunov}
%  Let $\Sigma = (\R^n, P, \Pow, F)$ be a switched system. A smooth
%   function $V\colon \R^n \times \R^n \rightarrow \Rplus$ is called  a
%   \emph{common $\delta$-GAS Lyapunov function for $\Sigma$} if there
%   exist $\K_\infty$ functions $\underline{\alpha}$, $\overline{\alpha}$
%   and $\kappa>0$ that make
%   the following hold for all $x, y \in \R^n$.
% \begin{align*}
% & \underline{\alpha}(\|x-y\|)\leq V(x, y) \leq \overline{\alpha}(\|x-y\|)
% %\label{eq:defValphacommon}
% \\
% & \frac{\partial V}{\partial x}(x, y)f_p(x) +
%  \frac{\partial V}{\partial y}(x, y)f_p(y) \leq -\kappa V(x, y) \quad\text{for all }p\in P
% %\label{eq:defVkappacommon}
% \end{align*}
% \end{mydef}

\begin{mythm}[\cite{GirardPT10}]
 \label{thm:commondeltaGAS}
 Let $\Sigma$ be a switched system. Assume that all continuous subsystems $\Sigma_p (p\in\Pow)$ have a $\delta$-GAS Lyapunov function $V$ in common (with the same $\kappa$).
Then, $V$ is called \emph{a common $\delta$-GAS Lyapunov function for $\Sigma$}, and $\Sigma$ is $\delta$-GUAS.
\qed
\end{mythm}

Another sufficient condition is the existence of \emph{multiple
$\delta$-GAS Lyapunov functions}, under an additional assumption on the
set of  switching signals.
The use of multiple Lyapunov functions for hybrid and switched systems
is first advocated in~\cite{branicky1998multiple}.
We let $\Signal_{\tau_d}(\Rplus, P)\subseteq\Signal(\Rplus, P)$ denote
the set of switching signals with a \emph{dwell time} $\tau_d > 0$, which means that the intervals between switching times are always longer than $\tau_d$.

We introduce the following notations.
Given a  switched system $\Sigma =
(\R^n, P=\{1, 2, \cdots, m\}, \Pow, F)$, % , recall that
 % $\Sigma_{p}$ denotes the switching-free subsystem where the mode is
 % fixed to $p$ (see~\S{\ref{sec:switchedSys}}). 
assume that, for each $p\in\{1,2,\dotsc,m\}$, we have a $\delta$-GAS Lyapunov function $V_p$ for the
 subsystem $\Sigma_{p}$. 
% single-mode subsystem $\Sigma_{p}$ of the switched system $\Sigma =
% (\R^n, P=\{1, 2, \cdots, p, \cdots, m\}, \Pow, F)$
% (see~\S{\ref{sec:switchedSys}}).
Then there % exist (sean)
exist
 a constant $\kappa_p\in\Rplus$ and two $\K_\infty$ functions $\underline{\alpha}_p$ and $\overline{\alpha}_p$ as in Def.~\ref{def:deltaGASLyapunov}.
% If $V_p$ has a $\delta$-GAS Lyapunov function for all $p\in P=\{1, 2,
% \cdots, m\}$, then
Let us now define
\begin{equation}\label{eq:constantsForMultipleLyapunov}
 \begin{aligned}
\begin{array}{ll}
  \underline{\alpha} \defeq \min(\underline{\alpha}_1, 
%\underline{\alpha}_2, 
\dotsc, \underline{\alpha}_m)\enspace, &
 \overline{\alpha} \defeq \max(\overline{\alpha}_1, 
%\overline{\alpha}_2, 
\dotsc, \overline{\alpha}_m)\enspace, \\
 \kappa \defeq \min(\kappa_1, 
%\kappa_2, 
\dotsc, \kappa_m)\enspace.
\end{array}
\end{aligned}
\end{equation}

\begin{mythm}[\cite{GirardPT10}]\label{thm:multipledeltaGAS}
 Let
 %$\Sigma$
 $\Sigma =
(\R^n, P, \Pow, F)$
 be a switched system. Assume that
 $ P=\{1, 2, \dotsc, m\}$, and that
 its set $\Pow$ of switching signals satisfies
 $\Pow\subseteq\Signal_{\tau_d}(\Rplus, P)$.
Assume further that, for
each $p\in P$, there exists a $\delta$-GAS Lyapunov function $V_p$ for the subsystem $\Sigma_{p}$.
We also assume that there exists $\mu\in\Rplus$ such that
$$V_p(x, y)\leq \mu V_{p'}(x, y) \quad\text{for all } x, y\in\R^n \text{ and } p, p'\in P.$$
If the dwell time $\tau_{d}$ satisfies $\tau_d > \frac{\log \mu}{\kappa}$, then $\Sigma_{\tau_d}$ is $\delta$-GUAS.
\qed
\end{mythm}

\section{Approximate Bisimulation for Delays I: Common Lyapunov Functions}
\label{sec:common}
 We have reviewed two witness notions for the incremental
stability notion of  $\delta$-GUAS: common and multiple $\delta$-GAS Lyapunov
functions.
%In~\cite{GirardPT10}, 
These two notions have been previously used mainly for discrete-state
abstraction of switched systems (see~\S{}\ref{sec:relatedWork}).

It is our main contribution to use the
same incremental stability assumptions to derive upper bounds for errors caused by switching delays. We focus on    periodic switching systems;
our translation of them to transition systems (Def.~\ref{def:TSigma}) plays an essential  role. 

The use of common $\delta$-GAS Lyapunov
functions is described in this section; the use of multiple  $\delta$-GAS Lyapunov
functions is in the next section.
The proofs in both sections are omitted.
See Appendix~\ref{sec:incrementing} for their proofs.

%In our technical development 
We will be using the following assumption.
\begin{myasm}[bounded intermode derivative]\label{asm:boundedIntermodeDerivative}
Let
  $\Sigma =
(\R^n, P, \Pow, F)$
 be a switched system, with $P=\{1,2,\dotsc, m\}$ and $F = \{f_1, f_2, \dotsc, f_m\}$ being the set of vector fields associated with each mode. 
We say a function $V\colon \R^n \times \R^n \rightarrow \Rplus$ has \emph{bounded intermode derivatives} if there exists a real number $\nu\geq 0$ such that, 
for any $p,p'\in P$ that are distinct ($p\neq p'$),
the inequality
\begin{align}
 \frac{\partial V}{\partial x}(x, y)f_{p}(x) + \frac{\partial V}{\partial y}(x, y)f_{p'}(y) \leq \nu\label{eq:additionalAssum}
\end{align}
holds for each $x,y\in \R^n$. 
%When the condition is satisfied we define $\nu\defeq\max_{p, p'}\nu_{p, p'}$. 
\end{myasm}

% for some $\nu_{p, p'}\geq 0$ for all $p, p'\in P$.

\begin{myrem}
Assumption~\ref{asm:boundedIntermodeDerivative} seems to be new: it is not assumed in the previous works on approximate bisimulation for switched systems% \red{, such as~\cite{GirardPT10}}
. 
Imposing the assumption on $\delta$-GAS Lyapunov functions, however, is not a severe restriction. In~\cite{GirardPT10} they make the assumption 
\begin{equation}\label{eq:GirardPTTriangleAssumption}
\exists \gamma\in\Rplus.\; \forall x, y, z\in\R^n.\; |V(x, y) - V(x, z)|\leq\gamma(\|y-z\|)
\end{equation}
 (we do not need this assumption in the current work). It is claimed in~\cite{GirardPT10} that~(\ref{eq:GirardPTTriangleAssumption}) is readily guaranteed if the dynamics of the switched system is confined to a compact set  $C\subseteq\R^n$, and if $V$ is class $\mathcal{C}^{1}$ in the domain $C$. We can use the same compactness argument to ensure Assumption~\ref{asm:boundedIntermodeDerivative}. 
% This was not assumed in the paper \cite{GirardPT10}.
% However, in \cite{GirardPT10} they assumed that
% $$\forall x, y, z\in\R^n, |V(x, y) - V(x, z)|\leq\gamma(\|y-z\|).$$
% This assumption was justified because if we consider the dynamics of a switched system in a compact set $C\subseteq\R^n, $ then it is sufficient to assume that $V$ is $\mathcal{C}^1$ in $C$.
% The same justification applies to our assumption (\ref{eq:additionalAssum}).
\end{myrem}

\begin{mydef}[the function $V'$] \label{def:VprimeFromV}
Let   $\Sigma =
(\R^n, P, \Pow, F)$
be a switched system, and let $V\colon \R^n \times \R^n \rightarrow \Rplus$ be a common $\delta$-GAS Lyapunov function for $\Sigma$. 

We define a function $V':(\R^n\times\Rplus\times P)\times (\R^n\times \Rplus\times P)\rightarrow \Rplus$ in the following manner:
\begin{align*}
& V'\bigl((x, t, p), (x', t', p')\bigr)\defeq\\
& \qquad\begin{cases}
 V\bigl(x, \x{x'}{t-t'}{p'}\bigr) & \text{ if }p = p' \text{ and }t\in[t', t'+\delta_{0}]\\
 \infty & \text{ otherwise.}
\end{cases}
\end{align*}
Recall that $\x{x'}{t-t'}{p'}$ is the state reached from $x'$ after time $t-t'$ following the vector field $f_{p'}$. 
\end{mydef}

Here is our main technical lemma. 
%The proof is omitted because it is easy once we have Lem.~\ref{lem:mainCommonIncrementing}.
\begin{mylem}\label{lem:mainCommon}
Let $\Sigma_{\tau} = (\R^n, P, \Pow_{\tau}, F)$ be a $\tau$-periodic switched system, and $\Sigma_{\tau, \delta_{0}} = (\R^n, P, \Pow_{\tau, \delta_{0}}, F)$ be a $\tau$-periodic switched system with delays within $\delta_{0}$. 
Assume that there exists
%Then existence of
 a common $\delta$-GAS Lyapunov function $V$ for $\Sigma_{\tau}$, and that $V$ satisfies the additional assumption in Assumption~\ref{asm:boundedIntermodeDerivative}. 

We consider a relation $R_\varepsilon\subseteq(\R^n\times\Rplus\times P)\times (\R^n\times\Rplus\times P)$ defined by
\begin{equation}
 (q, q')\in R_\varepsilon\;\overset{\mathrm{def.}}{\Longleftrightarrow}\; V'(q, q')\leq\underline\alpha(\varepsilon)\enspace.\label{eq:R} 
\end{equation}
Here $V'$ is from Def.~\ref{def:VprimeFromV}. 
If we fix $\varepsilon = \underline{\alpha}^{-1}\left(
\frac{
%\max_{p, p'}\nu_{p, p'}
\nu\delta_{0}}{1-e^{-\kappa(\tau-\delta_{0})}}
\right)$ 
where $\nu$ is from Assumption~\ref{asm:boundedIntermodeDerivative}, then, the relation $R_\varepsilon$ is an approximate bisimulation
between the transition systems $T(\Sigma_{\tau, \delta_{0}})$ and $T(\Sigma_\tau)$.
\qed

% Then it holds that the 
%  family $\{R_\varepsilon\}_{\varepsilon \geq 0}$ is an $g$-incrementing approximate bisimulation between 
% $%T_{\tau, \delta_{0}}
% T(\Sigma_{\tau, \delta_{0}})$ and 
% $%T_\tau
% T(\Sigma_\tau)$. 
%  If there exists a common $\delta$-GAS Lyapunov function $V$ for $\Sigma_{\tau}$ and the supplementary assumption in (\ref{eq:additionalAssum}) is satisfied, then
% the set of relations $R_\varepsilon\subseteq(\R^n\times\Rplus\times P)\times (\R^n\times\Rplus\times P)$ indexed by $\varepsilon\geq 0$ defined by
% $(q, q')\in R_\varepsilon\overset{\mathrm{def}}{\Leftrightarrow} V'(q, q')\leq\underline\alpha(\varepsilon)$
% is a set of relaxed approximate bisimulation relations between $T_{\tau, \delta_{0}}(\Sigma_{\tau, \delta_{0}})$ and $T_\tau(\Sigma_\tau)$ by taking the function $f$ by 
% $$f(\varepsilon) = \underline\alpha^{-1}\left(e^{-\kappa(\tau-\delta_{0})}\underline\alpha(\varepsilon) + \max_{p, p'}\nu_{p, p'}\delta_{0}\right).$$
\end{mylem}

In the following theorem, we compare the trajectories of $\Sigma_{\tau,\delta_{0}}$ and $\Sigma_{\tau}$ from the same initial state $x$.
It is a direct consequence from the previous lemma.
\begin{mythm}\label{thm:maincommon}
Assume the same assumptions as in Lem.~\ref{lem:mainCommon}. 
Let $\mathbf{p}_{\tau}$ be a $\tau$-periodic switching signal, and 
 $\mathbf{p}_{\tau,\delta_{0}}$  be the same signal but with  delays within $\delta_{0}$. That is,
for each $s\in\Rplus$,
\begin{displaymath}
{\small
 \begin{aligned}
  \mathbf{p}_{\tau,\delta_{0}}(s)
  =
   \begin{cases}
    \mathbf{p}_{\tau}(s) \text{ or } \mathbf{p}_{\tau}(s-\delta_{0})
    &\text{if $s\in \bigcup_{k\in\N, k\ge 1}[k\tau,k\tau+\delta_{0})$}
    \\
    \mathbf{p}_{\tau}(s) &
    \text{otherwise.}
 %    \text{if $s\in[0,\delta_{0})\cup\bigcup_{k\in\N}[k\tau+\delta_{0},(k+1)\tau)$}
   \end{cases}
 \end{aligned}
}
\end{displaymath}

% \begin{displaymath}
%   \mathbf{p}_{\tau,\delta_{0}}(s)
%   =
%    \begin{cases}
%     \mathbf{p}_{\tau}(s) \text{ or } \mathbf{p}_{\tau}(s-\delta_{0})
%     &\text{if $s\in \bigcup_{k\in\N, k\ge 1}[k\tau,k\tau+\delta_{0})$}
%     \\
%     \mathbf{p}_{\tau}(s) &
%     \text{otherwise.}
% %    \text{if $s\in[0,\delta_{0})\cup\bigcup_{k\in\N}[k\tau+\delta_{0},(k+1)\tau)$}
%    \end{cases}
% \end{displaymath} 

%(See Fig.~\ref{fig:withAndWithoutDelays} for illustration.)
 We have, for each $t\in\Rplus$, 
\begin{align*}
&\bigl\|\x{x}{t}{\mathbf{p}_{\tau,\delta_{0}}}
- \x{x}{t}{\mathbf{p}_\tau}
\bigr\|\;\leq\;
\underline{\alpha}^{-1}\left(
\frac{
%\max_{p, p'}\nu_{p, p'}
\nu\delta_{0}}{1-e^{-\kappa(\tau-\delta_{0})}}
\right)
\enspace.
\tag*{\myqed}
\end{align*}
\end{mythm}

Note that, for any desired precision $\varepsilon$, there always exists a small enough delay bound $\delta_{0}$ that achieves
the precision $\varepsilon$ (i.e.\ $\frac{
%\max_{p, p'}\nu_{p, p'}
\nu\delta_{0}}{1-e^{-\kappa(\tau-\delta_{0})}}
 \leq \varepsilon$).

\begin{myrem}\label{rem:overlineAlphaNotNecessary}
It turns out that the upper bound  $\overline{\alpha}$  of a 
$\delta$-GAS Lyapunov function $V$ (see~(\ref{eq:defValpha})) is not used
in the above results nor their proofs.
In \cite{GirardPT10}, % the upper bound
 $\overline{\alpha}$ is used to 
define the state space discretization parameter $\eta$ so that, for each initial state $q_1\in I_1$, there would be an approximately bisimilar initial state in $I_2$ and vice versa.
This is not necessary 
in our current setting where
%since our usage of approximate bisimulation is not the state space discretization and 
there is an obvious correspondence between the initial states.
That   $\overline{\alpha}$  is unnecessary  is also the case with the multiple Lyapunov function case in the next section.
\end{myrem}

\section{Approximate Bisimulation for Delays II: Multiple Lyapunov Functions}\label{sec:multiple}
We % \red{follow~\cite{GirardPT10} and} 
investigate the use of another witness for $\delta$-GUAS incremental stability---namely multiple  $\delta$-GAS Lyapunov
functions, see~\S{\ref{sec:incrStability}}---for bounding errors caused by switching delays. 

% , the authors extend their results in which a common Lyapunov function is given, to the case where a common Lyapunov function does not exist but multiple Lyapunov functions exist.
% In this section, we extend our result in a similar way.

The following is an analogue of Assumption~\ref{asm:boundedIntermodeDerivative}. 
\begin{myasm}\label{asm:boundedIntermodeDerivativeMultiple}
Let
  $\Sigma =
(\R^n, P, \Pow, F)$
 be a switched system with $P=\{1,2,\dotsc, m\}$. Let $V_{1},\dotsc,V_{m}\colon \R^n \times \R^n \rightarrow \Rplus$ be smooth functions. 
We say the functions
$V_{1},\dotsc,V_{m}$ 
% $V\colon \R^n \times \R^n \rightarrow \Rplus$ 
have \emph{bounded intermode derivatives} if
there exists a real number
$\nu'\geq 0$ 
such that,
for each $p,p'\in P$ that are distinct ($p\neq p'$),
the inequality
\begin{align}
 \frac{\partial V_{p'}}{\partial x}(x, y)f_{p}(x) + \frac{\partial V_{p'}}{\partial y}(x, y)f_{p'}(y) \leq \nu'\label{eq:additionalAssumMultiple}
\end{align}
holds for each $x,y\in \R^n$. (Note the occurrences of $p$ and $p'$.)
\end{myasm}

Under Assumption~\ref{asm:boundedIntermodeDerivativeMultiple} and the dwell-time assumption, namely, 
 $\tau-\delta_0 > \frac{\log\mu}{\kappa}$, we can establish an approximate bisimulation between the transition systems $T(\Sigma_{\tau, \delta_{0}})$ and $T(\Sigma_\tau)$.
% For this multiple Lyapunov function case, we slightly change the supplementary assumption introduced in (\ref{eq:additionalAssum}) into
% \begin{align}
%  \frac{\partial V_{p'}}{\partial x}(x, y)f_{p}(x) + \frac{\partial V_{p'}}{\partial y}(x, y)f_{p'}(y) \leq \nu'_{p, p'}\label{eq:additionalAssumMultiple}
% \end{align}

\begin{mylem}\label{lem:mainMultiple}
Let $\Sigma_{\tau} = (\R^n, P, \Pow_{\tau}, F)$ be a $\tau$-periodic switched system and $\Sigma_{\tau, \delta_{0}} = (\R^n, P, \Pow_{\tau, \delta_{0}}, F)$ be a $\tau$-periodic switched system with delays within $\delta_{0}$.
Assume that for each $p\in P$, there is a $\delta$-GAS Lyapunov function $V_p$ for the single-mode subsystem $\Sigma_{\tau, p}$.
We also assume Assumption~\ref{asm:boundedIntermodeDerivativeMultiple} for $V_{1},\dotsc, V_{m}$, and that there exists $\mu\in\Rplus$ such that
\begin{align}\label{eq:RelBetweenDifferentLyapunov}
V_p(x, y)\leq \mu V_{p'}(x, y) \;\text{for all } x, y\in\R^n \text{ and } p, p'\in P\enspace.
\end{align}
The last assumption is the same as in Thm.~\ref{thm:multipledeltaGAS}. 
Additionally, we assume the dwell-time assumption $\tau-\delta_0 > \frac{\log\mu}{\kappa}$.

We consider a relation $R_\varepsilon\subseteq(\R^n\times\Rplus\times P)\times (\R^n\times\Rplus\times P)$ defined by
\begin{equation}
 (q, q')\in R_\varepsilon\;\overset{\mathrm{def.}}{\Longleftrightarrow}\; V'(q, q')\leq\underline\alpha(\varepsilon)\enspace.\label{eq:R} 
\end{equation}
Here the function $V'$ is defined as follows, adapting Def.~\ref{def:VprimeFromV} to the current multiple Lyapunov function setting. 
\begin{align*}
& V'\bigl((x, t, p), (x', t', p')\bigr)\defeq\\
& \qquad\begin{cases}
 V_{p}\bigl(x, \x{x'}{t-t'}{p'}\bigr) & \text{ if }p = p' \text{ and }t\in[t', t'+\delta_{0}]\\
 \infty & \text{ otherwise.}
\end{cases}
\end{align*}

If we fix $\varepsilon = \underline{\alpha}^{-1}\left(
\frac{
%\max_{p, p'}\nu_{p, p'}
\nu\delta_{0}}{1-\mu e^{-\kappa(\tau-\delta_{0})}}
\right)$ 
where $\underline{\alpha}$ and  $\kappa$ are from~(\ref{eq:constantsForMultipleLyapunov}) and $\nu'$ is from Assumption~\ref{asm:boundedIntermodeDerivativeMultiple},
 then, the relation $R_\varepsilon$ is an approximate bisimulation
between the transition systems $T(\Sigma_{\tau, \delta_{0}})$ and $T(\Sigma_\tau)$.
\end{mylem}

% Using Lem.~\ref{lem:mainMultiple}, the following theorem can be proved in the same way as Thm.~\ref{thm:maincommon}.
The next result follows directly from  Lem.~\ref{lem:mainMultiple}. 
\begin{mythm}\label{thm:mainMultiple}
Assume the same assumptions as in Lem.~\ref{lem:mainMultiple}, and let 
$\mathbf{p}_{\tau}$ and
 $\mathbf{p}_{\tau,\delta_{0}}$ be those periodic switching signals, without and with delays, as in Thm.~\ref{thm:maincommon}. 
 If $\tau-\delta_0 > \frac{\log\mu}{\kappa}$, we have, for each $t\in\Rplus$, 
\begin{align*}
&\bigl\|\x{x}{t}{\mathbf{p}_{\tau,\delta_{0}}}
- \x{x}{t}{\mathbf{p}_\tau}
\bigr\|\leq
\underline{\alpha}^{-1}\left(\frac{
%\max_{p, p'}\nu_{p, p'}
\nu'\delta_{0}}{1-\mu e^{-\kappa(\tau-\delta_{0})}}
\right)
\enspace.
\tag*{\myqed}
\end{align*}
\end{mythm}

\section{Example}\label{sec:example}
%In this section we showcase how our methodology described so far works, using two examples.
%In this section
 We demonstrate 
%how our methodology works
our framework using the example of the boost DC-DC converter from~\cite{BeccutiPM05}.
It is a common example of  switched systems% \red{that is also used in \cite{GirardPT10}}
. For this example  we have a common $\delta$-GAS Lyapunov function $V$, and therefore we appeal to the results in~\S{\ref{sec:common}}.
% for an approximate bisimulation. 
We also % \red{combine our analysis with the analysis in~\cite{GirardPT10},}
demonstrate the control synthesis workflow in Fig.~\ref{fig:usageScenario}. 

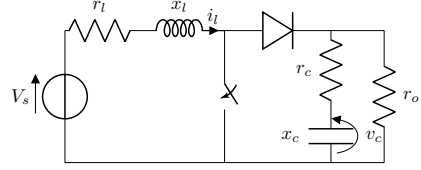
\begin{figure}[tbp]
\scalebox{.7}{ \begin{circuitikz}
 \draw
 (0, 0) to[V=$V_s$] (0, 2.5)
 to[R, l=$r_l$] (1.3, 2.5)
 to[L, l=$x_l$, i>=$i_l$] (3, 2.5)
 to[switch] (3, 0)
 -- (0, 0);

 \draw
 (3, 0) -- (5, 0)
 to[C, l=$x_c$, v>=$v_c$] (5, 1)
 to[R, l=$r_c$] (5, 2.5)
 -- (6, 2.5)
 to[R, l=$r_o$] (6, 0)
 -- (5, 0);

 \draw
 (3, 2.5) to[D] (5, 2.5);
 \end{circuitikz}}
\caption{The boost DC-DC converter circuit.}
\label{fig:boostDCDC}
\end{figure}

Another example of a water tank with nonlinear dynamics is presented in Appendix~\ref{subsec:watertank}.
 It has multiple $\delta$-GAS Lyapunov functions, and we use the results in~\S{\ref{sec:multiple}}.
\paragraph*{System Description}\label{para:DCDCSystem}
The system we consider is the boost DC-DC converter in Fig.~\ref{fig:boostDCDC}.
It is taken from~\cite{BeccutiPM05}; here we follow and extend the analysis in~\cite{GirardPT10}. The circuit includes a capacitor with capacitance $x_c$ and an inductor with inductance $x_l$.
The capacitor has the equivalent series resistance $r_c$, and the inductor has the internal resistance $r_l$.
The input voltage is $v_s$, and the resistance $r_o$ is the output load resistance.
The state 
$x(t)=
\begin{bmatrix}
 i_l(t)\\
 v_c(t)		 
\end{bmatrix}$
 of this system consists of the inductor current $i_l$ and the capacitor voltage $v_c$.

The dynamics of this system has two modes $\{\ON, \OFF\}$\footnote{In the formalization of~\S{\ref{sec:switchedSys}}, the set $P$ of modes is declared as $\{1, \cdots, m\}$. Here we instead use $P = \{\ON, \OFF\}$ for readability. % The same applies to the  water tank example in Appendix~{\ref{subsec:watertank}}.
},  depending on whether the switch in the circuit is on or off. By elementary circuit theory, the dynamics in each mode is modeled by
\begin{align*}
& \dot{x}(t) = A_p x(t) + b \quad\text{for }p\in\{\ON, \OFF\}\enspace,\text{ where}
\\
& A_{\ON} =
 \begin{bmatrix}
  -\frac{r_l}{x_l} & 0\\
  0 & -\frac{1}{x_c(r_o+r_c)}
 \end{bmatrix},
 \quad
 b =
 \begin{bmatrix}
  \frac{v_s}{x_l}\\
  0
 \end{bmatrix} \text{ and}\\
& A_{\OFF} =
 \begin{bmatrix}
  -\frac{r_l r_o + r_l r_c + r_o r_c}{x_l(r_o+r_c)} & -\frac{r_l r_o + r_l r_c + r_o r_c}{x_l(r_o+r_c)}\\
  \frac{r_o}{x_c(r_o+r_c)} & -\frac{1}{x_c(r_o+r_c)}
 \end{bmatrix}\enspace.
\end{align*}
We 
%follow the approach used in \cite{GirardPT10} and take
use the parameter values from \cite{BeccutiPM05}, that is,
$x_c =70$ p.u., $x_l =3$ p.u., $r_c = 0.005$ p.u., $r_l =0.05$ p.u., $r_o =1$ p.u. and $v_s =1$ p.u.
%\red{The same parameter values are used in~\cite{GirardPT10}.}

\paragraph*{Analysis}
Following \cite{GirardPT10}, we rescale the second variable of the system and redefine the state 
$x(t)=
\begin{bmatrix}
 i_l(t)\\
 5v_c(t)
\end{bmatrix}$ for better numerical conditioning.
The ODEs are updated accordingly.

It is shown in~\cite{GirardPT10} that
the dynamics in each mode is $\delta$-GAS. They share a common $\delta$-GAS Lyapunov function
$V(x, y) = \sqrt{(x-y)^T M (x-y)}$,
with
$M=
\begin{bmatrix}
 1.0224 & 0.0084\\
 0.0084 & 1.0031 
\end{bmatrix}
$. 
The common Lyapunov function $V$ has $\underline{\alpha}(s) = s, \overline{\alpha}(s) = 1.0127s$ and $\kappa = 0.014$.
This common Lyapunov function is discovered in~\cite{GirardPT10} via SDP optimization; we use the same function as an ingredient for our approximate bisimulation. 

Our ultimate goal is to synthesize a switching signal that keeps the dynamics in a safe region
% $\mathcal{S}\defeq [1.3, 1.7]\times[5.7, 5.8]$. 
$\mathcal{S}\defeq [1.3, 1.7]\times[5.7, 5.8]
%\enspace. 
$.
We shall follow the two-step workflow in Fig.~\ref{fig:usageScenario}. 

Let us first use Thm.~\ref{thm:maincommon} and derive a bound $\varepsilon_{1}$ for errors caused by switching delays. We set the
switching period $\tau = 0.5$ % \red{(this is the same as in~\cite{GirardPT10})}, 
 and the maximum delay $\delta_{0} = \frac{\tau}{1000}$. On top of the analysis in~\cite{GirardPT10}, we have to verify the condition we additionally impose (namely Assumption~\ref{asm:boundedIntermodeDerivative}). Let us now assume that the dynamics stays in the safe region $\mathcal{S}=[1.3, 1.7]\times[5.7, 5.8]$---this assumption will be eventually discharged when we synthesize a safe controller. Then it is not hard to see that 
$\nu = 0.41$ satisfies the inequality~(\ref{eq:additionalAssum}).
By Thm.~\ref{thm:maincommon}, we obtain that the error between $\Sigma_{\tau, \delta_{0}}$ 
(the boost DC-DC converter with delays)
and $\Sigma_\tau$ (the one without delays) is bounded by $\varepsilon_{1} = 0.0294176$.

% Our scenario is that we would like to control the switch so that the state will stay in $[1.3, 1.7]\times[5.7, 5.8]$.
% For this region, we checked that $\nu_{\ON, \OFF} = 0.41$ satisfies (\ref{eq:additionalAssum}) for both $(p, p')=(\ON, \OFF)$ and $(\OFF, \ON)$.
% By applying Thm.~\ref{thm:maincommon}, we obtain that the error between $\Sigma_{\tau, \delta_{0}}$ and $\Sigma_\tau$ is bounded by $\varepsilon = 0.0294176$ from above.

We sketch how we can combine the above analysis with the analysis in~\cite{GirardPT10}, in the way prescribed in Fig.~\ref{fig:usageScenario}. In~\cite{GirardPT10} they use the same Lyapunov function as above to derive a discrete symbolic model $T_{\tau}^{\text{symb}}$ and establish an approximate bisimulation between $T(\Sigma_{\tau})$ and the symbolic model. Their symbolic model  $T_{\tau}^{\text{symb}}$ can be constructed so that any desired error bound $\varepsilon_{2}$ is guaranteed (a smaller $\varepsilon_{2}$ calls for a finer grid for discretization and hence a bigger symbolic model). 

Now we employ an algorithm from supervisory control in~\cite{RamadgeW87}, and synthesize a set of safe switching signals that confine the dynamics of $T_{\tau}^{\text{symb}}$ to a shrunk safe region
% \begin{align*}
% &  \mathcal{S}_{\varepsilon_1+\varepsilon_2}\defeq
%  [1.3+(\varepsilon_1+\varepsilon_2), 1.7-(\varepsilon_1+\varepsilon_2)]
% \\&\qquad\qquad\times[5.7+(\varepsilon_1+\varepsilon_2), 5.8-(\varepsilon_1+\varepsilon_2)]\enspace.
% \end{align*}
\begin{math}
  \mathcal{S}_{\varepsilon_1+\varepsilon_2}\defeq
  [1.3+(\varepsilon_1+\varepsilon_2), 1.7-(\varepsilon_1+\varepsilon_2)]
\times[5.7+(\varepsilon_1+\varepsilon_2), 5.8-(\varepsilon_1+\varepsilon_2)]
\end{math}.
Let $\mathbf{p}$ be any such safe switching signal.
 By the second approximate bisimulation in Fig.~\ref{fig:usageScenario}, the signal  $\mathbf{p}$ is guaranteed to keep the dynamics of $\Sigma_{\tau}$ in the region
\begin{math}
  \mathcal{S}_{\varepsilon_1}\defeq
 [1.3+\varepsilon_1, 1.7-\varepsilon_1]
\times[5.7+\varepsilon_1, 5.8-\varepsilon_1]
\end{math}. Finally, the first approximate bisimulation  in Fig.~\ref{fig:usageScenario} guarantees that the signal  $\mathbf{p}$ keeps the dynamics of $\Sigma_{\tau,\delta_{0}}$, the system with switching delays, in $\mathcal{S}$. 
\begin{myrem}\label{rem:shrunkSafeSpace}
On the choice of a safe region used in control synthesis for the symbolic model $T_{\tau}^{\text{symb}}$, our current choice  $\mathcal{S}_{\varepsilon_1+\varepsilon_2}\subsetneq \mathcal{S}$ 
%(see~(\ref{eq:shrunkSafeRegion})) 
is more conservative than the choice in~\cite{GirardPT10}, where they in fact expand (rather than shrink) the original safe region $\mathcal{S}$. We believe our conservative choice is required in the current workflow (Fig.~\ref{fig:usageScenario}) where two approximation steps are totally separated. Tighter integration of the two steps can lead to relaxation of this conservative choice. 
\end{myrem}

\section{Related Work}\label{sec:relatedWork}
The notion of approximate bisimulation is first introduced
in~\cite{Girard2007}. Use of incremental stability as a source of approximate
bisimulations is advocated in~\cite{PolaGT08}. This useful technique has found its applications in a
variety of system classes as well as in a variety of problems. A notable application is  discretization of continuous state
spaces so that discrete verification/synthesis techniques can be
employed. The original framework in~\cite{PolaGT08} has seen its
extension to switched systems (\cite{GirardPT10}), systems with
disturbance (\cite{PolaT09}), and so on. A comprehensive framework where
discrete control synthesis is integrated is presented in~\cite{Girard10}; the
works discussed so far are nicely summarized in the overview
paper by~\cite{Girard2011}. 

A line of works relevant to ours addresses the issue of
time delays (\cite{PolaPBT10,PolaPB10}). The work by~\cite{PolaPBT10} deals
with fixed time delays and the one by~\cite{PolaPB10} considers unknown
time delays. The goal of these works, which is different from ours, is to
construct a comprehensive symbolic (discretized) model that encompasses
all possible delays and switching signals. In particular, possible
delays are thought of as disturbances (i.e.\ demonic/adversarial
nondeterminism) and consequently they use \emph{alternating} approximate
bisimulations. The main technical gadget in doing so is spline-based
finitary approximation of continuous-time signals.

Towards control synthesis for switched
systems with unknown switching delays, our workflow
(Fig.~\ref{fig:usageScenario}) is two-step while a workflow based on~\cite{PolaPB10} is one-step. 
% the difference between the
% approach via~\cite{PolaPB10} and ours (Fig.~\ref{fig:usageScenario}) can
% be understood as one-step vs.\ two-step. 
The latter  works as follows.  The results
in~\cite{PolaPB10} yields a symbolic model for a switched system with delays; it is given by a two-player finite-state game $\mathcal{G}$ where
angelic moves switching signals and demonic moves are time delays.
By solving the game  $\mathcal{G}$ (e.g.\ by the algorithm in~\cite{Jurdzinski00}) one
obtains a control strategy. It seems that our two-step workflow has an
advantage in complexity: by collecting the spline-based approximations
of all possible delays and switching signals, the game  $\mathcal{G}$
in~\cite{PolaPB10} tends to have a big number of transitions. 
% In comparison, our workflow---that separates
% time delays from control synthesis, 
% Fig.~\ref{fig:usageScenario}---seems to have an advantage in
% complexity: in control synthesis we do not need to cope with a vast
% variety of (spline-based approximations of) unknown time delays. 
It has
to be noted, however, that the workflow following~\cite{PolaPB10}
applies to a greater variety of systems (than switched systems) and
a resulting control strategy can be more fine-grained (reacting to delays, while our controller always assumes the worst
time delays).

Another related work that refers to time delays is~\cite{LIU20161}.
The biggest difference between \cite{LIU20161} and our work is that their framework is based on the invariance assumption of atomic formulas ($\delta$ in the paper): as long as errors due to delays do not exceed $\delta$, the system satisfies the same set of LTL formulas. In contrast, our results bound distances between trajectories; the use of such bounds is not restricted to satisfaction of LTL specifications.

The works~\cite{BorriPD12,ZamaniMKA17} study symbolic abstraction of networked control systems, taking into accounts issues including time delays. The main difference from the current work is that their delays are assumed to be always multiples ($0,\tau,2\tau,\dotsc$) of the period $\tau$; this assumption is enforced in their framework by system components called the \emph{zero-order hold (ZOH)}. Their game-based frameworks are based on alternating approximate bisimulations, much like in~\cite{PolaPB10}, but the above assumption leads to simpler games for control synthesis. We note that our current setting---where delays are within a fixed bound $\delta_{0}$ with $\delta_{0}<\tau$---is  outside the scope of~\cite{BorriPD12,ZamaniMKA17}.

A recent line of works by~\cite{KhatibGD16,KhatibGD17} tackles the challenge of time
delays too. They take \emph{timing contracts} as specifications; and study
a verification problem (\cite{KhatibGD16}), and a scheduling problem under
the   single-processor  multiple-task
setting (\cite{KhatibGD17}). A crucial difference from the current work is
that they assume linear dynamics, while we can deal with nonlinear
dynamics (under the assumption of incremental stability).

The works by~\cite{AbbasMF14,AbbasF15} present an
alternative notion of
proximity between trajectories called  \emph{$(\tau,
\varepsilon)$-closeness}. Its intuition is that, in establishing that the distance between two trajectories is within $\varepsilon$, it is allowed to shift time by $\tau$. 
Adapting our current results to this notion of proximity does not seem
hard, by
modifying the definitions in Def.~\ref{def:pseudoquasimetricBetweenTrajectories} and~\ref{def:VprimeFromV}.

\section{Conclusions and Future Work}\label{sec:conclusion}
In this paper% \red{, based on the results in~\cite{GirardPT10}}
, we introduced
an approximate bisimulation framework and provided upper bounds for
errors that arise from switching delays in periodic switched
systems. Our focus on switched systems allows us to use the same
incremental stability notion ($\delta$-GUAS) as in~\cite{GirardPT10} as an ingredient
for an approximate bisimulation. This is an advantage in the control synthesis
workflow (Fig.~\ref{fig:usageScenario})  for
switched systems with delays, in which we separate two concerns of time
delays and discretization of state spaces. 

Adaptation of the current framework to  $(\tau,
\varepsilon)$-closeness in~\cite{AbbasMF14,AbbasF15} is  imminent future
work. 
% While the proximity notion of  $(\tau,
% \varepsilon)$-closeness is more permissive  than the
% one we use here (namely the Euclidean distance),
% for a certain class of specifications 
% %, see e.g.\Thm.~\ref{thm:maincommon}), 
% %it is shown in~\cite{AbbasMF14} that 
%  $(\tau,
% \varepsilon)$-closeness is sufficient~\cite{AbbasMF14}.
%  for ensuring those specifications which are
% expressed in a certain variant of metric temporal logic. 
As we discussed
in~\S{\ref{sec:relatedWork}}, this adaptation will not be hard.

Extending the current results to a wider class of systems beyond
periodic switched ones is  a direction that we shall
pursue. In particular, we are interested in disturbances
and the consequent use of alternating approximate
bisimulation introduced in~\cite{PolaT09}. Such an extension should be carefully
devised so that the two-step workflow in Fig.~\ref{fig:usageScenario}
remain valid. For example, once  the
controller synthesized based on the symbolic model
$T_{\tau}^{\text{symb}}$ involves sensing, it is unlikely that the controller achieves
precision $\varepsilon_{1}+\varepsilon_{2}$ in the system
$\Sigma_{\tau,\delta_{0}}$ with delays. In such a setting, errors in each of the two
approximation steps in Fig.~\ref{fig:usageScenario} can amplify each
other, resulting in an overall error bound that is much worse than
$\varepsilon_{1}+\varepsilon_{2}$. A related topic of future work is suggested in Rem.~\ref{rem:shrunkSafeSpace}.

\auxproof{
Introducing feedback control to our current framework. In such settings it is less clear how the combination of our current framework and the control synthesis framework by symbolic models, as sketched in Fig.~\ref{fig:usageScenario}, would work. The synthesized controller should necessarily depend on the sampling of current states, and the sampling results can differ between $T(\Sigma_{\tau,\delta_{0}})$ and $T(\Sigma_{\tau})$ due to switching delays. Our future work is to establish a unified framework for synthesis of feedback controllers in presence of switching (or rather actuation) delays, where our current results are more tightly integrated to symbolic model-based synthesis frameworks like those in~\cite{Girard12,MajumdarMS16}. Since feedback control will be needed only in presence of disturbance, in the envisaged uniform framework we would be using \emph{alternating} approximate bisimulation~\cite{PolaT09}.

}

%\bibliography{library}

\appendix

\section{Relaxation of approximate bisimulation for time-varying error bounds}
\label{sec:incrementing}

In this appendix, we introduce a relaxed notion of approximate bisimulation to obtain
 a tighter error bound that can grow in time. 
We present the proofs for the lemmas and the theorems in this appendix.
We do not show the omitted proofs for the results stated in \S{}\ref{sec:common} and \S{}\ref{sec:multiple} themselves, because they are direct consequences from the results in this appendix.

Our relaxation of approximate bisimulation allows errors to potentially grow after each transition, in a manner regulated by some increasing function $g\colon \Rplus\to\Rplus$. 
\begin{mydef}[$g$-incrementing approximate bisimulation]\label{def:incrementingApproxBisim}
%sean
  % Let $T_i = (Q_i, L, \Ra{}{i}, O, H_i, I_i)$ ($i=1$, $2$) be two premetric transition systems with premetric $d$; they share the same sets of actions $L$ and outputs $O$. Let $\varepsilon\in\Rplus$ be a \emph{precision}, and $f\colon \Rplus\to\Rplus$ be an increasing function. 
Let $T_i = (Q_i, L, \Ra{}{i}, O, H_i, I_i)$ ($i=1$, $2$) be two premetric transition systems with premetric $d$; they share the same sets of actions $L$ and outputs $O$. Let $\varepsilon\in\Rplus$ be a \emph{precision}, and $g\colon \Rplus\to\Rplus$ be an increasing function.

%Given a precision $\varepsilon\in\Rplus$, 

A family $\{R_\varepsilon\}_{\varepsilon \geq 0}$ of relations $R_\varepsilon\subseteq Q_1\times Q_2$ indexed by $\varepsilon\geq 0$ is called a \emph{$g$-incrementing approximate bisimulation} between $T_{1}$ and $T_{2}$ 
% a set of \emph{relaxed approximate bisimulation relations between $T_1$ and $T_2$} 
if the following  conditions hold for all $\varepsilon\geq 0$ and for all $(q_1, q_2)\in R_\varepsilon$:
\begin{itemize}
 \item $d(H_1(q_1), H_2(q_2))\leq\varepsilon$; and
 \item there exists a function $g$ such that
 \begin{itemize}
  \item  $\forall q_1\Ra{l}{1}q'_1, \exists q_2\Ra{l}{2}q'_2 \text{ s.t. } (q'_1, q'_2)\in R_{g(\varepsilon)}$;
\\ and
  \item $\forall q_2\Ra{l}{2}q'_2, \exists q_1\Ra{l}{1}q'_1 \text{ s.t. } (q'_1, q'_2)\in R_{g(\varepsilon)}$.
 \end{itemize}
\end{itemize}
\end{mydef}

We can establish a $g$-incrementing approximate bisimulation between the transition systems $T(\Sigma_{\tau, \delta_{0}})$ and $T(\Sigma_\tau)$.

\begin{mylem}\label{lem:mainCommonIncrementing}
Let $\Sigma_{\tau} = (\R^n, P, \Pow_{\tau}, F)$ be a $\tau$-periodic switched system, and $\Sigma_{\tau, \delta_{0}} = (\R^n, P, \Pow_{\tau, \delta_{0}}, F)$ be a $\tau$-periodic switched system with delays within $\delta_{0}$. 
Assume that there exists
%Then existence of
 a common $\delta$-GAS Lyapunov function $V$ for $\Sigma_{\tau}$, and that $V$ satisfies the additional assumption in Assumption~\ref{asm:boundedIntermodeDerivative}. 
Then, for a suitable $g$, there exists a $g$-incrementing approximate bisimulation
 $\{R_\varepsilon\}_{\varepsilon \geq 0}$  between the transition systems
$%T_{\tau, \delta_{0}}
T(\Sigma_{\tau, \delta_{0}})$ and 
$%T_\tau
T(\Sigma_\tau)$. 

Specifically, we define a function $g$ by
$$g(\varepsilon) \defeq \underline\alpha^{-1}\left(e^{-\kappa(\tau-\delta_{0})}\underline\alpha(\varepsilon) + 
%\max_{p, p'}\nu_{p, p'}
\nu\delta_{0}\right)\enspace,$$
where $\nu$ is from Assumption~\ref{asm:boundedIntermodeDerivative}. 
For each $\varepsilon \geq 0$, we define a relation 
$R_\varepsilon\subseteq(\R^n\times\Rplus\times P)\times (\R^n\times\Rplus\times P)$ 
by
\begin{equation}
 (q, q')\in R_\varepsilon\;\overset{\mathrm{def.}}{\Longleftrightarrow}\; V'(q, q')\leq\underline\alpha(\varepsilon)\enspace.\label{eq:R} 
\end{equation}
Here $V'$ is from Def.~\ref{def:VprimeFromV}. 
% Then it holds that the 
%  family $\{R_\varepsilon\}_{\varepsilon \geq 0}$ is an $g$-incrementing approximate bisimulation between 
% $%T_{\tau, \delta_{0}}
% T(\Sigma_{\tau, \delta_{0}})$ and 
% $%T_\tau
% T(\Sigma_\tau)$. 
%  If there exists a common $\delta$-GAS Lyapunov function $V$ for $\Sigma_{\tau}$ and the supplementary assumption in (\ref{eq:additionalAssum}) is satisfied, then
% the set of relations $R_\varepsilon\subseteq(\R^n\times\Rplus\times P)\times (\R^n\times\Rplus\times P)$ indexed by $\varepsilon\geq 0$ defined by
% $(q, q')\in R_\varepsilon\overset{\mathrm{def}}{\Leftrightarrow} V'(q, q')\leq\underline\alpha(\varepsilon)$
% is a set of relaxed approximate bisimulation relations between $T_{\tau, \delta_{0}}(\Sigma_{\tau, \delta_{0}})$ and $T_\tau(\Sigma_\tau)$ by taking the function $f$ by 
% $$f(\varepsilon) = \underline\alpha^{-1}\left(e^{-\kappa(\tau-\delta_{0})}\underline\alpha(\varepsilon) + \max_{p, p'}\nu_{p, p'}\delta_{0}\right).$$
\end{mylem}

\begin{myproof}%\marginpar{Ichiro to Kengo: I haven't yet checked the proof carefully. Will do so soon.}
It is obvious from the monotonicity of $\underline{\alpha}$ that $(q_{\tau, \delta_{0}}, q_{\tau})\in R_\varepsilon$ implies $d(q_{\tau, \delta_{0}}, q_{\tau})\leq\varepsilon$.

For $q_{\tau, \delta_{0}} = (x_{\tau, \delta_{0}}, t_{\tau, \delta_{0}}, p_{\tau, \delta_{0}})$ and $q_{\tau} = (x_{\tau}, t_{\tau}, p_{\tau})$, we assume that $q_{\tau, \delta_{0}}\Ra{l}{\tau, \delta_{0}}q_{\tau, \delta_{0}}' = (x'_{\tau, \delta_{0}}, t'_{\tau, \delta_{0}}, p'_{\tau, \delta_{0}})$ and $(q_{\tau, \delta_{0}}, q_{\tau})\in R_\varepsilon$ for some $\varepsilon$.
Our goal is to show that there exists $q_{\tau}' = (x'_{\tau}, t'_{\tau}, p'_{\tau})$ such that $q_{\tau}\Ra{l}{\tau}q_{\tau}'$ and $(q_{\tau, \delta_{0}}', q_{\tau}')\in R_{g(\varepsilon)}$.

 By the definition (\ref{eq:R}) of the relation $R_\varepsilon$ and the construction of the transition systems $T_{\tau, \delta_{0}}(\Sigma_{\tau, \delta_{0}})$ and $T_{\tau}(\Sigma_{\tau})$, it is easy to see that $p_{\tau, \delta_{0}} = p_{\tau}$, $t_{\tau} = k\tau$, $t_{\tau, \delta_{0}} \in [k\tau, k\tau+\delta_{0}]$ and $t'_{\tau, \delta_{0}} \in [(k+1)\tau, (k+1)\tau+\delta_{0}]$ for some $k\in\N$.
Then we define $q_{\tau}'$ by $q_{\tau}' = (\x{x_\tau}{\tau}{p_{\tau}}, (k+1)\tau, p'_{\tau, \delta_{0}})$ to obtain $q_{\tau}\Ra{l}{\tau}q_{\tau}'$.

Now we show $(q_{\tau, \delta_{0}}', q_{\tau}')\in R_{g(\varepsilon)}$ for this $q_{\tau}'$.
By the assumption $(q_{\tau, \delta_{0}}, q_{\tau})\in R_\varepsilon$, we have $V'(q_{\tau, \delta_{0}}, q_{\tau})\leq \underline\alpha(\varepsilon)$, which means
\begin{align}
V(x_{\tau, \delta_{0}}, \x{x_\tau}{t_{\tau, \delta_{0}} - t_{\tau}}{p_\tau})\leq \underline\alpha(\varepsilon).\label{eq:vk}
\end{align}
Note that this equation refers to the states of the two systems at $t=t_{\tau, \delta_{0}}$.
%Our goal is to bound $V'(q'_{\tau, \delta_{0}}, q'_{\tau})$ from above.
When time progresses for $t_{\tau}'-t_{\tau, \delta_{0}}$ with mode $p_{\tau, \delta_{0}} = p_\tau$ for both systems from $t=t_{\tau, \delta_{0}}$, we have
\begin{align}
& V(\x{x_{\tau, \delta_{0}}}{t_{\tau}'-t_{\tau, \delta_{0}}}{p_{\tau}}, \x{\x{x_\tau}{t_{\tau, \delta_{0}} - t_{\tau}}{p_\tau}}{t_{\tau}'-t_{\tau, \delta_{0}}}{p_{\tau}})\nonumber\\
& \leq
e^{-\kappa(t_{\tau}'-t_{\tau, \delta_{0}})}V(x_{\tau, \delta_{0}}, \x{x_\tau}{t_{\tau, \delta_{0}} - t_{\tau}}{p_\tau})\nonumber \\
&\leq
e^{-\kappa(\tau - \delta_{0})}V(x_{\tau, \delta_{0}}, \x{x_\tau}{t_{\tau, \delta_{0}} - t_{\tau}}{p_\tau}).\label{vk+1tau}
\end{align}
Note that this equation refers to the states of the two systems at $t=t_{\tau}'$.
When time progresses for $t_{\tau, \delta_{0}}'-t_{\tau}'$ with mode $p_{\tau, \delta_{0}} (= p_{\tau})$ for $T_{\tau, \delta_{0}}(\Sigma_{\tau, \delta_{0}})$ and with mode $p'_{\tau}$ for $T_{\tau}(\Sigma_{\tau}))$ from $t=t_{\tau}'$, we have
\begin{align}
& \quad V'(q'_{\tau, \delta_{0}}, q'_{\tau})\nonumber\\
&=  V(\x{\x{x_{\tau, \delta_{0}}}{t_{\tau}'-t_{\tau, \delta_{0}}}{p_{\tau}}}{t_{\tau, \delta_{0}}'-t_{\tau}'}{p_{\tau}},\nonumber\\
&\qquad \x{\x{\x{x_\tau}{t_{\tau, \delta_{0}} - t_{\tau}}{p_\tau}}{t_{\tau}'-t_{\tau, \delta_{0}}}{p_{\tau}}}{t_{\tau, \delta_{0}}'-t_{\tau}'}{p'_{\tau}})\nonumber\\
&\leq
e^{-\kappa(\tau - \delta_{0})}V(x_{\tau, \delta_{0}}, \x{x_\tau}{t_{\tau, \delta_{0}} - t_{\tau}}{p_\tau}) + \nu% _{p_\tau, p'_{\tau}}
 (t_{\tau, \delta_{0}}'-t_{\tau}')\nonumber\\
&\leq
e^{-\kappa(\tau - \delta_{0})}V(x_{\tau, \delta_{0}}, \x{x_\tau}{t_{\tau, \delta_{0}} - t_{\tau}}{p_\tau}) + \nu% _{p_\tau, p'_{\tau}}
 \delta_{0}\nonumber\\
&=  e^{-\kappa(\tau-\delta_{0})} V'(q_{\tau, \delta_{0}}, q_{\tau}) + % \max_{p, p'}\nu_{p, p'}
\nu \delta_{0}\nonumber\\
&\leq e^{-\kappa(\tau-\delta_{0})} \underline{\alpha}(\varepsilon) + % \max_{p, p'}\nu_{p, p'}
\nu \delta_{0}\nonumber\\
&= \underline{\alpha}(g(\varepsilon)).\label{eq:vk+1taudelta}
\end{align}
In the RHS of the first inequality in (\ref{eq:vk+1taudelta}),
the first term is from (\ref{vk+1tau}), which bounds the distance of the states at time $t'_\tau$. The second term overapproximates the effect of the behavior from $t'_\tau$ to $t'_{\tau,\delta_0}$, when the two systems are operated in different modes. 

Thus we have $(q_{\tau, \delta_{0}}', q_{\tau}')\in R_{g(\varepsilon)}$.
The converse simulation condition is shown similarly.
%  The proof of 
% \begin{math} 
%  \forall q_\tau\Ra{l}{\tau}q'_\tau, \exists q_{\tau, \delta_{0}}\Ra{l}{\tau, \delta_{0}}q'_{\tau, \delta_{0}} \text{ s.t. } (q'_{\tau, \delta_{0}}, q'_\tau)\in R_{g(\varepsilon)}
% \end{math}
% is similar.
 \myqed
% \begin{align*}
% V'(q'_{\tau, \delta_{0}}, q'_{\tau})\leq \underline\alpha \left(\underline\alpha^{-1} \left(e^{-\kappa(\tau-\delta_{0})} \underline\alpha(\varepsilon) + % \max_{p, p'}\nu_{p, p'}
% \nu \delta_{0}\right)\right).
% \end{align*}
\end{myproof}

\begin{mythm}\label{thm:maincommonIncrementing}
Assume the same assumptions as in Lem.~\ref{lem:mainCommonIncrementing}. 
Let $\mathbf{p}_{\tau}$ be a $\tau$-periodic switching signal, and 
 $\mathbf{p}_{\tau,\delta_{0}}$  be the same signal but with  delays within $\delta_{0}$. That is,
for each $s\in\Rplus$,
\begin{displaymath}
  \mathbf{p}_{\tau,\delta_{0}}(s)
  =
   \begin{cases}
    \mathbf{p}_{\tau}(s) \text{ or } \mathbf{p}_{\tau}(s-\delta_{0})
    &\text{if $s\in \bigcup_{k\in\N, k\ge 1}[k\tau,k\tau+\delta_{0})$}
    \\
    \mathbf{p}_{\tau}(s) &
    \text{otherwise.}
%    \text{if $s\in[0,\delta_{0})\cup\bigcup_{k\in\N}[k\tau+\delta_{0},(k+1)\tau)$}
   \end{cases}
\end{displaymath} 
%(See Fig.~\ref{fig:withAndWithoutDelays} for illustration.)

 We have, for each $k\in \N$ and $t\in [k\tau,(k+1)\tau)$, 
\begin{align*}
&\bigl\|\x{x}{t}{\mathbf{p}_{\tau,\delta_{0}}}
- \x{x}{t}{\mathbf{p}_\tau}
\bigr\|\;\leq\\
&\quad\underline\alpha^{-1}\left(\frac{
%\max_{p, p'}\nu_{p, p'}
\nu
\delta_{0}}{1-e^{-\kappa(\tau-\delta_{0})}}+e^{-\kappa(\tau-\delta_{0})k}\left(-\frac{
%\max_{p, p'}\nu_{p, p'}
\nu
\delta_{0}}{1-e^{-\kappa(\tau-\delta_{0})}}\right)\right)
\enspace.
\end{align*}
\end{mythm}

\begin{myproof}
Lem.~\ref{lem:mainCommonIncrementing} serves as a recurrence relation with respect to the number $k$ of switching.
By solving it with the initial condition of $d(q_{\tau, \delta_{0}, 0}, q_{\tau, 0}) = 0$, 
we obtain the result of
% $$V'(q_{\tau, \delta_{0}}, q_{\tau})\leq \frac{% \max_{p, p'}\nu_{p, p'}
% \nu \delta_{0}}{1-e^{-\kappa(\tau-\delta_{0})}}+e^{-\kappa(\tau-\delta_{0})k}\left(-\frac{% \max_{p, p'}\nu_{p, p'}
% \nu \delta_{0}}{1-e^{-\kappa(\tau-\delta_{0})}}\right),$$
% and therefore
\begin{align*} 
 &d(q_{\tau, \delta_{0}}, q_{\tau})\\
\leq &\ g^k(0) \\
= &\ \underline\alpha^{-1}\left(\frac{% \max_{p, p'}\nu_{p, p'}
\nu \delta_{0}}{1-e^{-\kappa(\tau-\delta_{0})}}+e^{-\kappa(\tau-\delta_{0})k}\left(-\frac{% \max_{p, p'}\nu_{p, p'}
\nu \delta_{0}}{1-e^{-\kappa(\tau-\delta_{0})}}\right)\right),
\end{align*}
for all states $q_{\tau, \delta_{0}}$ and $q_{\tau}$ that can be reached via a same sequence of actions of length $k$.

Since the definition of the premetric $d$ is as Def.~\ref{def:pseudoquasimetricBetweenTrajectories}, this result only refers to the error between two systems at the switching time in $[k\tau, (k+1)\tau)$.
It is easy, however, to have that this $\quad\underline\alpha^{-1}\left(\frac{% \max_{p, p'}\nu_{p, p'}
\nu\delta_{0}}{1-e^{-\kappa(\tau-\delta_{0})}}+e^{-\kappa(\tau-\delta_{0})k}\left(-\frac{% \max_{p, p'}\nu_{p, p'}
\nu \delta_{0}}{1-e^{-\kappa(\tau-\delta_{0})}}\right)\right)$ is actually an upper bound of 
$\bigl\|\x{x}{t}{\mathbf{p}_{\tau,\delta_{0}}}
- \x{x}{t}{\mathbf{p}_\tau}
\bigr\|$, for all
$t\in [k\tau,(k+1)\tau)$, from $\kappa \geq 0$ and $\nu \geq 0$.
\qed
\end{myproof}

Note that the bound in Thm.~\ref{thm:maincommonIncrementing} can grow over time (i.e.\ over the number $k$ of switching), while the one in Thm.~\ref{thm:maincommon} is a conservative bound that is time-invariant.

The following lemma is for the multiple Lyapunov functions.
Note that we do not need the dwell-time assumption, which is necessary to obtain the approximate bisimulation in Lem.~\ref{lem:mainMultiple}.

\begin{mylem}\label{lem:mainMultipleIncrementing}
Let $\Sigma_{\tau} = (\R^n, P, \Pow_{\tau}, F)$ be a $\tau$-periodic switched system and $\Sigma_{\tau, \delta_{0}} = (\R^n, P, \Pow_{\tau, \delta_{0}}, F)$ be a $\tau$-periodic switched system with delays within $\delta_{0}$.
Assume that for each $p\in P$, there is a $\delta$-GAS Lyapunov function $V_p$ for the single-mode subsystem $\Sigma_{\tau, p}$.
We also assume Assumption~\ref{asm:boundedIntermodeDerivativeMultiple} for $V_{1},\dotsc, V_{m}$, and that there exists $\mu\in\Rplus$ such that
\begin{align}\label{eq:RelBetweenDifferentLyapunov}
V_p(x, y)\leq \mu V_{p'}(x, y) \;\text{for all } x, y\in\R^n \text{ and } p, p'\in P\enspace.
\end{align}

Then, for a suitable $g$, there exists a $g$-incrementing approximate bisimulation
 $\{R_\varepsilon\}_{\varepsilon \geq 0}$  between the transition systems
$%T_{\tau, \delta_{0}}
T(\Sigma_{\tau, \delta_{0}})$ and 
$%T_\tau
T(\Sigma_\tau)$. 

Specifically, we define $g$ by
$$g(\varepsilon) \defeq 
\underline\alpha^{-1}\left(\mu e^{-\kappa(\tau-\delta_{0})}\underline\alpha(\varepsilon) + 
%\max_{p, p'}\nu'_{p, p'}
\nu'\delta_{0}\right)
\enspace,
$$
where $\underline{\alpha}$ and  $\kappa$ are from~(\ref{eq:constantsForMultipleLyapunov}) and $\nu'$ is from Assumption~\ref{asm:boundedIntermodeDerivativeMultiple}. 
For each $\varepsilon \geq 0$, we define a relation 
$R_\varepsilon\subseteq(\R^n\times\Rplus\times P)\times (\R^n\times\Rplus\times P)$ 
by
\begin{math}
 (q, q')\in R_\varepsilon\;\overset{\mathrm{def.}}{\Longleftrightarrow}\; V'(q, q')\leq\underline\alpha(\varepsilon)
%\enspace. 
\end{math}.
Here the definition of the function $V'$ is the same as Lem.~\ref{lem:mainMultiple}.
\end{mylem}

\begin{myproof}
The proof of this lemma is almost the same as that of Lem.~\ref{lem:mainCommon}.
The only difference is, after we derive 
\begin{align*}
&\quad V_p(\x{x_{\tau, \delta_{0}}}{t_{\tau}'-t_{\tau, \delta_{0}}}{p_{\tau}},\\
&\qquad \x{\x{x_\tau}{t_{\tau, \delta_{0}} - t_{\tau}}{p_\tau}}{t_{\tau}'-t_{\tau, \delta_{0}}}{p_{\tau}})\\
&\leq e^{-\kappa(\tau - \delta_{0})}V_p(x_{\tau, \delta_{0}}, \x{x_\tau}{t_{\tau, \delta_{0}} - t_{\tau}}{p_\tau}),
\end{align*}
 which is the counterpart of the inequality (\ref{vk+1tau}), we derive
% the inequality
\begin{align*}
&\quad V_{p'}(\x{x_{\tau, \delta_{0}}}{t_{\tau}'-t_{\tau, \delta_{0}}}{p_{\tau}},\\
&\qquad \x{\x{x_\tau}{t_{\tau, \delta_{0}} - t_{\tau}}{p_\tau}}{t_{\tau}'-t_{\tau, \delta_{0}}}{p_{\tau}})\\
&\leq \mu e^{-\kappa(\tau - \delta_{0})}V_p(x_{\tau, \delta_{0}}, \x{x_\tau}{t_{\tau, \delta_{0}} - t_{\tau}}{p_\tau})
\end{align*}
 by using (\ref{eq:RelBetweenDifferentLyapunov}).
Then,  similarly to (\ref{eq:vk+1taudelta}), we obtain the inequality
\begin{align*}
 & \quad V_{p'}(\x{\x{x_{\tau, \delta_{0}}}{t_{\tau}'-t_{\tau, \delta_{0}}}{p_{\tau}}}{t_{\tau, \delta_{0}}'-t_{\tau}'}{p_{\tau}},\nonumber\\
&\qquad \x{\x{\x{x_\tau}{t_{\tau, \delta_{0}} - t_{\tau}}{p_\tau}}{t_{\tau}'-t_{\tau, \delta_{0}}}{p_{\tau}}}{t_{\tau, \delta_{0}}'-t_{\tau}'}{p'_{\tau}})\nonumber\\
&\leq
\mu e^{-\kappa(\tau - \delta_{0})}V(x_{\tau, \delta_{0}}, \x{x_\tau}{t_{\tau, \delta_{0}} - t_{\tau}}{p_\tau}) + \nu'\delta_{0}.
\end{align*}
The rest of the proof is straightforward.
 \qed
\end{myproof}

The next result follows directly from Lem.~\ref{lem:mainMultipleIncrementing}. 
\begin{mythm}\label{thm:mainMultiple}
Assume the same assumptions as in Lem.~\ref{lem:mainMultipleIncrementing}, and let 
$\mathbf{p}_{\tau}$ and
 $\mathbf{p}_{\tau,\delta_{0}}$ be those periodic switching signals, without and with delays, as in Thm.~\ref{thm:maincommon}. 
We have, for each $k\in \N$ and $t\in [k\tau,(k+1)\tau)$, 
\begin{align*}
&\bigl\|\x{x}{t}{\mathbf{p}_{\tau,\delta_{0}}}
- \x{x}{t}{\mathbf{p}_\tau}
\bigr\|\;\leq\\
&\quad
\underline\alpha^{-1}\left(\frac{
%\max_{p, p'}\nu_{p, p'}
\nu'
\delta_{0}}{1-\mu e^{-\kappa(\tau-\delta_{0})}}+\mu e^{-\kappa(\tau-\delta_{0})k}\left(-\frac{
%\max_{p, p'}\nu_{p, p'}
\nu'
\delta_{0}}{1-\mu e^{-\kappa(\tau-\delta_{0})}}\right)\right)
% \underline\alpha^{-1}\left(\frac{
% %\max_{p, p'}\nu_{p, p'}
% \nu
% \delta_{0}}{1-e^{-\kappa(\tau-\delta_{0})}}+e^{-\kappa(\tau-\delta_{0})k}\left(-\frac{
% %\max_{p, p'}\nu_{p, p'}
% \nu
% \delta_{0}}{1-e^{-\kappa(\tau-\delta_{0})}}\right)\right)
\enspace.
\end{align*}
\end{mythm}

\section{Another Example of Nonlinear Water tank}
\label{subsec:watertank}
\paragraph*{System Description}
This example  demonstrates  our framework's applicability to nonlinear dynamics. 
The water tank we consider is equipped with a drain and a valve. The system has two modes.
When the switch is off, the drain is open and the valve is closed, causing the water level to decrease. When the switch is on, the drain is closed and the valve aperture is set according to the water level.
% \begin{enumerate}
%  \item When the switch is off, the drain is open and the valve is closed. Thus there is no inflow and the water level decreases.
%  \item When the switch is on, the drain is closed and the valve aperture is set according to the water level.
% \end{enumerate}
We assume that dynamics of the water level $x$ is modeled by:
\begin{align}
 \dot{x}=
\begin{cases}
f_{\OFF}(x)\defeq -a\sqrt{x} \quad\text{when the switch is off,}\\
f_{\ON}(x)\defeq b(c-x) \quad\text{when the switch is on.}
\end{cases}
\end{align}
The behavior for the mode $\OFF$ is a well-known  water level behavior, found e.g.\ in the MATLAB\textregistered/Simulink\textregistered\ example (\cite{WatertankSimulink}).
The water leaves at a rate that is proportional to $\sqrt{x}$.
%The parameter $a$ is determined by the shape of the tank.
%\marginpar{There is a digital controller that}
The behavior of the mode $\ON$ is a natural one when the valve aperture is governed by a float, as found in many toilet tanks. 
%  inspired by the behavior of toilet tanks.
% In a toilet tank, there is a float in the tank and it automatically changes the aperture of the valve according to the position of the float, i.e., the water level $x$.
% The parameter $c$ is the height of the float where the valve gets completely closed.
% When the water level $x$ is lower than $c$, the valve is partially open and the water level increases.
% if the water tank is empty, the valve is totally open and the other parameter $b$ is to defined by the inflow rate at that time.

Let us set the three parameters  $a=\frac{1}{5}, b=\frac{1}{10}$ and $c=11$.
Our scenario is that we would like to control the switch so that the water level should stay in $[1,10]$.
We assume there are switching delays within $\delta_{0} = 0.1$ seconds.
We fix the switching period $\tau$ to be 10 seconds.

\paragraph*{Analysis}
The dynamics of each mode  has a $\delta$-GAS Lyapunov function defined by
\begin{align*}
 V_{\OFF}(x, y)\defeq|e^{\sqrt{x}}-e^{\sqrt{y}}|
  \quad\text{and}\quad
 V_{\ON}(x, y)\defeq|\sqrt{6}(x-y)|\enspace.
\end{align*}
We obtain the following characteristics for these two $\delta$-GAS Lyapunov functions in the safe region $[1,10]$:
% \begin{align*}
% & \underline\alpha_{\OFF}(s)=s,\quad
%  \underline\alpha_{\ON}(s)=\sqrt{6}s, \quad
%  \kappa_{\OFF} = \kappa_{\ON} = \frac{1}{10},\\
% & \mu = \frac{2\sqrt{6}}{3},\quad
%  \nu'_{\OFF} = 2.74\quad\text{and}\quad
%  \nu'_{\ON}= 2.94.
% \end{align*}
$\underline\alpha_{\OFF}(s)=s$,
$\underline\alpha_{\ON}(s)=\sqrt{6}s$,
$\kappa_{\OFF} = \kappa_{\ON} = \frac{1}{10}$,
$\mu = \frac{2\sqrt{6}}{3}$, and
$\nu' = 2.94$. 
% Thus we have
% $\underline\alpha(s)=s$, $\kappa = 1/10$, $\mu = \frac{2\sqrt{6}}{3}$ and $\nu' = 2.94$.

% Each of them satisfies the following inequalities for all $x, y$ in the target region $1\leq x, y \leq 10$.
% \begin{align*}
% & \underline\alpha_{\OFF}(\|x-y\|)\defeq\|x-y\|\leq V_{\OFF}(x, y)\\
% &\frac{\partial V_{\OFF}}{\partial x}(x, y)f_{\OFF}(x) +
%  \frac{\partial V_{\OFF}}{\partial y}(x, y)f_{\OFF}(y) \leq -\frac{1}{10} V_{\OFF}(x, y) \\
% & \underline\alpha_{\ON}(\|x-y\|)\defeq\sqrt{6}\|x-y\|\leq V_{\ON}(x, y)\\
% &\frac{\partial V_{\ON}}{\partial x}(x, y)f_{\ON}(x) +
%  \frac{\partial V_{\ON}}{\partial y}(x, y)f_{\ON}(y) \leq -\frac{1}{10} V_{\ON}(x, y) \\
% &V_{\OFF}(x, y)\leq \frac{2\sqrt{6}}{3}V_{\ON}(x, y)\\
% &V_{\ON}(x, y)\leq \frac{2\sqrt{6}}{3}V_{\OFF}(x, y)\\
% &\frac{\partial V_{\ON}}{\partial x}(x, y)f_{\OFF}(x) + \frac{\partial V_{\ON}}{\partial y}(x, y)f_{\ON}(y) \leq \nu'_{\OFF, \ON} \defeq 2.94\\
% &\frac{\partial V_{\OFF}}{\partial x}(x, y)f_{\ON}(x) + \frac{\partial V_{\OFF}}{\partial y}(x, y)f_{\OFF}(y) \leq \nu'_{\ON, \OFF} \defeq 2.74
% \end{align*}
% Thus we have
% $\underline\alpha(s) = s, \kappa = 1/10,
% \mu = \frac{2\sqrt{6}}{3}\text{ and }\nu' = 2.94.$
% Note that we do not use $\overline{\alpha}$ as discussed in Rem.~\ref{rem:overlineAlphaNotNecessary}.

By  Thm.~\ref{thm:mainMultiple} we obtain that the error between $\Sigma_{\tau, \delta_{0}}$ and $\Sigma_\tau$ is bounded by $\varepsilon = 0.747678$.
Much like in the previous example, if we design a controller for
% the periodic system
 $\Sigma_\tau$ so that
% the water level
 $x$ stays within $ [1+\varepsilon, 10-\varepsilon]$,
%by using the methodology in \cite{Girard10}, for example, 
then the same controller will keep the water level $x$ of the delayed system $\Sigma_{\tau, \delta_{0}}$ within $[1,10]$.

\begin{myrem}
Our Lyapunov functions for this example are not smooth at $x=y$.
Therefore, in (\ref{eq:defVkappa}) and (\ref{eq:additionalAssumMultiple}), the partial derivatives of $V_p$ are undefined.
This nonsmoothness can be dealt with using \emph{Clarke's nonsmooth analysis}, which nicely accommodates set-valued generalized derivatives.
More specifically, our functions are \emph{Clarke regular} and locally Lipschitz, and we can derive similar results as those in~\cite{BacciottiC99}.
\end{myrem}

\end{document}